\documentclass[aps, pra, reprint, floatfix, superscriptaddress, showkeys, nofootinbib, fleqn]{revtex4-2}

\pdfoutput=1

\usepackage{amsmath, amsfonts, amssymb, amsthm, bm, bbm, nccmath}

\usepackage[colorlinks=true, linkcolor=blue, citecolor=purple, urlcolor=blue]{hyperref}

\usepackage[activate={true,nocompatibility},final,tracking=true,kerning=true,spacing=true,factor=1100]{microtype}
\microtypecontext{spacing=nonfrench}

\usepackage{lmodern}
\usepackage[T1]{fontenc}
\usepackage[utf8]{inputenc}
\usepackage[american]{babel}

\usepackage{orcidlink}
\usepackage{titlesec}
\usepackage{nccmath}
\usepackage{tikz-cd}
\usepackage{tikz}
\usepackage{tikz-3dplot}

\usepackage{graphicx}
\usepackage[shortlabels]{enumitem}

\usepackage[left=1.5cm, right=1.5cm, top=1.7cm, bottom=1.7cm]{geometry}

\theoremstyle{remark}
\newtheorem{rmk}{Remark}
\newtheorem{ex}{Example}

\newcommand{\HH}{\mathcal{H}}
\newcommand{\KK}{\mathcal{K}}
\newcommand{\DD}{\mathcal{D}}
\newcommand{\WW}{\mathcal{W}}

\newcommand{\UU}{\mathrm{U}}

\newcommand{\uu}{\mathfrak{u}}

\renewcommand{\AA}{\mathcal{A}}

\newcommand{\T}{\mathrm{T}}
\newcommand{\V}{\mathrm{V}}
\renewcommand{\H}{\mathrm{H}}
\newcommand{\tr}{\operatorname{tr}}
\renewcommand{\d}{\mathrm{d}}
\newcommand{\dt}{\mathrm{d}t}
\newcommand{\ds}{\mathrm{d}s}
\newcommand{\bra}[1]{\langle #1|}
\newcommand{\ket}[1]{|#1\rangle}
\newcommand{\braket}[2]{\langle #1 | #2 \rangle}
\newcommand{\ketbra}[2]{|#1 \rangle\langle #2|}

\newcommand{\1}{\mathbbm{1}}
\newcommand{\length}{\mathcal{L}}
\newcommand{\lengthFR}{\mathcal{L}_{\textrm{\tiny FR}}}
\newcommand{\lengthFS}{\mathcal{L}_{\textrm{\tiny FS}}}
\newcommand{\energy}{\mathcal{E}}
\newcommand{\energyFR}{\mathcal{E}_{\textrm{\tiny FR}}}
\newcommand{\energyFS}{\mathcal{E}_{\textrm{\tiny FS}}}
\newcommand{\isobound}{\operatorname{iHB}}

\newcommand{\bfp}{\boldsymbol{p}}
\newcommand{\bfm}{\boldsymbol{m}}
\newcommand{\bfPi}{\boldsymbol{\Pi}}
\newcommand{\bfL}{\boldsymbol{\Lambda}}
\newcommand{\bfa}{\boldsymbol{\alpha}}

\newcommand{\thetageo}{\theta_{\textrm{geo}}}

\newcommand{\supp}{\operatorname{supp}}
\newcommand{\inco}{\textrm{\tiny in}}
\newcommand{\co}{\textrm{\tiny co}}
\newcommand{\gFR}{g_{\textrm{\tiny FR}}}   
\newcommand{\Ad}{\operatorname{Ad}}

\begin{document}
\title{Isoholonomic inequalities and speed limits for cyclic quantum systems}
\author{Ole S{\"o}nnerborn\,\orcidlink{0000-0002-1726-4892}\,}
\email{ole.sonnerborn@kau.se}
\affiliation{Department of Mathematics and Computer Science, Karlstad University, 651 88 Karlstad, Sweden}
\affiliation{Department of Physics, Stockholm University, 106 91 Stockholm, Sweden}

\date{\today}

\begin{abstract}
\noindent
Quantum speed limits set fundamental lower bounds on the time required for a quantum system to evolve between states. Traditional bounds, such as those by Mandelstam--Tamm and Margolus--Levitin, rely on state distinguishability and become trivial for cyclic evolutions where the initial and final states coincide. In this work, we explore an alternative approach based on isoholonomic inequalities, which bound the length of closed trajectories in the state space in terms of their holonomy. Building on a gauge-theoretic framework for mixed-state geometric phases, we extend the concept of isoholonomic inequalities to closed curves of isospectral and isodegenerate density operators. This allows us to derive a new quantum speed limit that remains nontrivial for cyclic evolutions. Our results reveal deep connections between the temporal behavior of cyclic quantum systems and holonomy.
\end{abstract}

\maketitle

\titleformat{\section}[block]{\bfseries\large}{\Roman{section}}{0.7em}{}
\titleformat{\subsection}[block]{\bfseries\normalsize}{\Roman{section}.\Alph{subsection}}{0.7em}{}
\titleformat{\subsubsection}[block]{\bfseries\small}{\Roman{section}.\Alph{subsection}.\arabic{subsubsection}}{0.4em}{}

\titlespacing\section{0pt}{1.2em}{0.8em}

\titlespacing\subsection{0pt}{1em}{0.5em}

\titlespacing\subsubsection{0pt}{0.7em}{0.3em}

\section{Introduction}
\label{sec: Introduction}
\noindent
A quantum speed limit is a fundamental lower bound on the minimum time required for a quantum system to evolve from one state to another. 

Most quantum speed limits, including the classical ones attributed to Mandelstam and Tamm~\cite{MaTa1991, AnAh1990, AnHe2014, HoAlSo2022} and Margolus and Levitin~\cite{MaLe1998, GiLlMa2003, HoSo2023a, HoSo2023b}, involve the fidelity or another measure of distinguishability between the initial and final states~\cite{Fr2016, DeCa2017}. This characteristic typically renders such limits trivial for cyclic evolutions, where the initial and final states coincide.

Isoholonomic inequalities, introduced by Montgomery~\cite{Mo1990}, offer a powerful tool for estimating the return time of cyclic processes. These inequalities provide lower bounds on the length of closed curves, such as the trajectory of the state of a periodic quantum system, in terms of the curve's holonomy. For example, the Fubini--Study length of a closed curve of pure states is bounded below by a quantity known as the isoholonomic bound. This bound is determined solely by the geometric phase, that is, the phase of the Aharonov--Anandan holonomy associated with the curve. From this isoholonomic inequality, it follows that the time required for a unitary evolving system to return to its initial state is at least the ratio of the isoholonomic bound to the system's time-averaged energy uncertainty~\cite{HoSo2023c}.

Isoholonomic inequalities have also been applied in the context of holonomic quantum computation. In this setting, quantum gates are implemented by driving the computational space through a closed loop within a larger Hilbert space, either adiabatically or nonadiabatically. The resulting gate corresponds to the holonomy associated with this cyclic evolution~\cite{ZaRa1999, ZhKyFiKwSjTo2023}. Isoholonomic inequalities have been applied to establish fundamental lower bounds on the execution time of these gates, thereby linking geometric constraints to computational performance~\cite{So2024a, So2025a, So2025b}.

In this paper, we derive isoholonomic inequalities and corresponding quantum speed limits for systems that undergo cyclic evolution through isodegenerate mixed states. To achieve this, we introduce the concept of length and define holonomy for closed curves within these states. Specifically, we show that the gauge-theoretic framework developed in Refs.~\cite{AnHe2015, An2018}, which underlies the geometric phase for mixed states introduced by Sj\"oqvist~\emph{et al.}~\cite{SjPaEkAnErOiVe2000} and Tong~\emph{et al.}~\cite{ToSjKwOh2004}, provides a natural geometric setting for formulating these isoholonomic inequalities and their associated speed limits.

The paper is organized as follows. Section~\ref{sec: Isoholonomic inequality and speed limit for cyclic systems in pure states} revisits the isoholonomic inequality and the associated quantum speed limit for systems evolving cyclically through pure states, as derived in Ref.~\cite{HoSo2023c}. Section~\ref{sec: Geometric phase for systems evolving through isodegenerate states} discusses the geometric phase for mixed isodegenerate states introduced in Refs.~\cite{SjPaEkAnErOiVe2000, ToSjKwOh2004}. Section~\ref{sec: Gauge theory for isospectral and isodegenerate states} presents the gauge-theoretic framework developed in Refs.~\cite{AnHe2015, An2018}. Section~\ref{sec: Isoholonomic inequalities} derives isoholonomic inequalities for closed curves of isospectral and isodegenerate states. Section~\ref{sec: Speed limit for the return time of unitarily evolving cyclic systems} establishes a quantum speed limit for the return time of cyclic unitarily evolving systems. Finally, Section~\ref{sec: Tightness of the isoholonomic inequality for isospectral dynamics} presents a tightness result.

\section{Isoholonomic inequality and speed limit for cyclic systems in pure states}%
\label{sec: Isoholonomic inequality and speed limit for cyclic systems in pure states}
\noindent
Consider a quantum system evolving through a piecewise smooth family of pure states $\rho_t$ that returns to the initial state at a time $\tau>0$. The isoholonomic inequality provides a lower bound---the isoholonomic bound---on the Fubini--Study length of this closed trajectory of the state~\cite{HoSo2023c}:
\begin{equation}
    \lengthFS[\rho_t] 
    \geq \isobound[\rho_t].
\label{eq: isoholonomic inequality for pure states}
\end{equation}
The Fubini--Study length quantifies the geometric distance traversed by the state and is defined as
\begin{equation}
    \lengthFS[\rho_t] 
    = \int_{0}^{\tau} \sqrt{ \frac{1}{2} \tr\big(\dot\rho_t^2\big) } \,\dt.
\end{equation}
The isoholonomic bound depends only on the representative $\theta $ of the  Aharonov--Anandan geometric phase of $\rho_t$ in the interval $[0,2\pi)$ and is given by
\begin{equation}
	\isobound[\rho_t] 
	= \sqrt{\theta(2\pi - \theta)}.
\end{equation}
Recall that the Aharonov--Anandan geometric phase is defined as the equivalence class modulo $2\pi$ of the phase of the Aharonov--Anandan holonomy associated with the curve $\rho_t$, which is defined as
\begin{equation}
    \Gamma[\rho_t]
    = \braket{\psi_0}{\psi_\tau}\exp\Big( - \int_{0}^{\tau} \braket{\psi_t}{\dot{\psi}_t}\,\dt \Big),
    \label{eq: Aharonov--Anandan holonomy}
\end{equation}
where $\ket{\psi_t}$ is any smooth curve of unit vectors satisfying $\rho_t = \ketbra{\psi_t}{\psi_t}$; see Ref.~\cite{AhAn1987}. The holonomy is a phase factor of purely geometric origin accumulated during the evolution, and the quantity $\theta$ appearing in the isoholonomic bound is its phase, taken in the interval $[0,2\pi)$.

If the system evolves unitarily, so that $\rho_t = U_t \rho U_t^\dagger$, with $U_t$ being the propagator generated by a Hamiltonian $H_t$, the Fubini--Study speed coincides with the instantaneous energy uncertainty (we adopt units in which $\hbar=1$):
\begin{equation}
    \sqrt{ \frac{1}{2} \tr\big(\dot\rho_t^2\big) } 
    = \Delta H_t,\quad \Delta^2 H_t=\tr\big(\rho_t H_t^2\big) - \tr\big(\rho_tH_t\big)^2.
\end{equation}
Letting $\Delta E$ denote the average energy uncertainty over the evolution time interval, the Fubini--Study length of the path $\rho_t$ can be expressed as $\lengthFS[\rho_t] = \tau \Delta E$. Inserting this relation into the isoholonomic inequality \eqref{eq: isoholonomic inequality for pure states}, we arrive at a quantum speed limit of Mandelstam--Tamm type that applies to cyclic evolutions \cite{HoSo2023c}:
\begin{equation}
    \tau \geq \frac{ \sqrt{\theta(2\pi - \theta)} }{\Delta E}.
    \label{eq: the QSL for a qubit}
\end{equation}
Unlike the classic Mandelstam--Tamm bound~\cite{MaTa1991, AnAh1990, AnHe2014, HoAlSo2022}, which relates the time it takes for a system to evolve between two states to the Fubini--Study distance between them, this bound provides a nontrivial lower limit on the time $\tau$ at which the system returns to its initial state.

\begin{ex}
A qubit in a pure state evolving unitarily under a time-independent Hamiltonian exhibits periodic dynamics. Reference~\cite{HoSo2023c} shows that the period saturates the inequality in Eq.~\eqref{eq: the QSL for a qubit}. Thus, if $\tau$ denotes the system's period and $\theta$ the Aharonov--Anandan geometric phase associated with the closed curve traced by the state over one period, then equality holds in Eq.~\eqref{eq: the QSL for a qubit}.
\label{rmk: pure}
\end{ex}

The isoholonomic inequality \eqref{eq: isoholonomic inequality for pure states} plays a central role in the derivation of the main result of this paper. A streamlined version of the proof from Ref.~\cite{HoSo2023c} is therefore included in Appendix \ref{app: The isoholonomic inequality for systems in pure states}.

\begin{rmk}
In this paper, a \emph{curve} of vectors, states, or linear maps refers to a continuous, piecewise smooth one-parameter family of such objects. The parameter, denoted by $t$ and called time, is assumed to range over the interval $[0,\tau]$, where $\tau>0$ is finite but otherwise unspecified. At points where the curve is not smooth, we assume the existence of one-sided derivatives.
\label{rmk: one-parameter family}
\end{rmk}

\section{Geometric phase for systems evolving through isodegenerate mixed states}%
\label{sec: Geometric phase for systems evolving through isodegenerate states}
\noindent
Consider a curve of density operators $\rho_t$ that share a spectrum of nondegenerate positive eigenvalues $p_1,p_2,\dots,p_l$.\footnote{The density operators have the same rank. If the rank is not full, $0$ is a common (possibly degenerate) eigenvalue. For reasons explained in Section~\ref{sec: Gauge theory for isospectral and isodegenerate states}, we exclude this eigenvalue from the spectrum.} Each density operator then admits a unique spectral decomposition of the form
\begin{equation}
    \rho_t = \sum_{j=1}^l p_j \ketbra{\psi_{j;t}}{\psi_{j;t}},
    \label{eq: nondeg spec dec}
\end{equation}
where $\ketbra{\psi_{j;t}}{\psi_{j;t}}$ is the orthogonal projection onto the one-dimensional eigenspace of $\rho_t$ associated with eigenvalue $p_j$. While these projectors are uniquely determined by $\rho_t$, each eigenvector $\ket{\psi_{j;t}}$ is defined only up to a phase factor.

For any choice of initial eigenvectors $\ket{\psi_{j;0}}$, there exist unique curves of instantaneous eigenvectors that satisfy the Aharonov--Anandan horizontality condition~\cite{AhAn1987}:
\begin{equation}
    \braket{\psi_{j;t}}{\dot{\psi}_{j;t}} = 0. 
    \label{eq: parallel} 
\end{equation} 
Following Sj\"{o}qvist~\emph{et al.}~\cite{SjPaEkAnErOiVe2000}, we refer to a spectral decomposition of $\rho_t$ in which each eigenvector evolves horizontally as a parallel-transported decomposition, and we define the geometric phase of $\rho_t$ as 
\begin{equation}
    \thetageo[\rho_t] = 
    \arg \sum_{j=1}^l p_j \braket{\psi_{j;0}}{\psi_{j;\tau}}. 
    \label{eq: ind geo phase} 
\end{equation}

When the common eigenvalues of the $\rho_t$s are degenerate, one must generalize the horizontality condition in Eq.~\eqref{eq: parallel} to account for the higher-dimensional eigenspaces~\cite{ToSjKwOh2004}: Suppose that $p_j$ has degeneracy $m_j$. Then $\rho_t$ admits a spectral decomposition of the form
\begin{equation}
    \rho_t = \sum_{j=1}^l \sum_{a=1}^{m_j} p_{j} \ketbra{\psi_{ja;t}}{\psi_{ja;t}},
\end{equation}
where for each $j$ and each $t$, the $\ket{\psi_{ja;t}}$s are normalized and pairwise orthogonal eigenvectors of $\rho_t$ spanning the eigenspace corresponding to the eigenvalue $p_j$. 
The natural generalization of the parallel-transport condition in this setting requires that for $j = 1,2, \dots, l$, the velocity vectors ${\ket{\dot\psi_{ja;t}}}$ lie orthogonal to the corresponding eigenspace. That is, the spectral decomposition is parallel transported if
\begin{equation} 
    \braket{\psi_{ja;t}}{\dot{\psi}_{jb;t}} = 0,
    \quad a,b=1,2,\dots, m_j. 
\label{eq: degenerate parallel transport condition}
\end{equation}
For a curve of density operators with a parallel-transported spectral decomposition, we define the geometric phase as
\begin{equation}
    \thetageo[\rho_t] 
    = \arg \sum_{j=1}^l \sum_{a=1}^{m_j} p_{j} \braket{\psi_{ja;0}}{\psi_{ja;\tau}}.
    \label{eq: deg spec dec}
\end{equation}

The parallel-transport condition~\eqref{eq: degenerate parallel transport condition} remains applicable even when the eigenvalues of the density operator change over time, provided that their multiplicities remain constant throughout the evolution~\cite{ToSjKwOh2004}. Let $p_{1;t},p_{2;t},\dots,p_{l;t}$ be the time-dependent eigenvalue spectrum of $\rho_t$, and let $m_1, m_2, \dots, m_l$ be the corresponding fixed multiplicities. In this setting, $\rho_t$ admits a spectral decomposition 
\begin{equation}
    \rho_t = \sum_{j=1}^l \sum_{a=1}^{m_j} p_{j;t} \ketbra{\psi_{ja;t}}{\psi_{ja;t}}
    \label{eq: spectral decomposition}
\end{equation}
with orthonormal eigenvectors $\ket{\psi_{ja;t}}$ that satisfy the parallel-transport condition~\eqref{eq: degenerate parallel transport condition}. The geometric phase associated with such a parallel-transported decomposition is defined as
\begin{equation}
    \thetageo[\rho_t]
    = \arg \sum_{j=1}^l \sum_{a=1}^{m_j} \sqrt{p_{j;0}p_{j;\tau}} \braket{\psi_{ja;0}}{\psi_{ja;\tau}}.
    \label{eq: clgp}
\end{equation}
This expression generalizes the earlier definitions, reducing to \eqref{eq: deg spec dec} when the spectrum is time independent and further simplifying to \eqref{eq: ind geo phase} in the case of a non\-degenerate spectrum.

\section{Gauge theories for isospectral and isodegenerate states}
\label{sec: Gauge theory for isospectral and isodegenerate states}
\noindent
In this section we review the gauge-theoretic framework developed in Refs.~\cite{AnHe2015, An2018}, which underlies the geometric phases introduced by Sj\"oqvist~\emph{et al.}~\cite{So2025b} and Tong~\emph{et al.}~\cite{ToSjKwOh2004}. We apply this framework in Section~\ref{sec: Isoholonomic inequalities} to generalize the isoholonomic inequality~\eqref{eq: isoholonomic inequality for pure states} to mixed states. We begin by introducing notation for the spectral structures relevant to mixed states.

A density operator is fully characterized by its set of distinct positive eigenvalues $\bfp = (p_1, p_2, \dots, p_l)$,
which we assume are arranged in descending order, and the corresponding family of orthogonal projection operators $\bfPi = (\Pi_1, \Pi_2, \dots, \Pi_l)$,
where $\Pi_j$ projects onto the eigenspace associated with the eigenvalue $p_j$. We refer to $\bfp$ as the eigenvalue spectrum and to $\bfPi$ as the eigenprojector spectrum of the density operator. If the density operator is not of full rank, $0$ is also an eigenvalue. However, for technical reasons, we exclude $0$ from $\bfp$ and omit the corresponding projection operator from $\bfPi$.

When an eigenprojector spectrum is specified without an accompanying eigenvalue spectrum, it is implicitly understood that any associated density operator possesses a set of positive eigenvalues ordered in descending magnitude, and that the $j$th projector corresponds to the eigenspace of the $j$th largest eigenvalue.

The rank of each projection operator $\Pi_j$ equals the multiplicity $m_j$ of the corresponding eigenvalue $p_j$. We refer to the sequence $\bfm = (m_1, m_2, \dots, m_l)$
as the degeneracy spectrum of the density operator. As before, no entry of $\bfm$ accounts for the multiplicity of a possible zero eigenvalue. Consequently, if a density operator has degeneracy spectrum $\bfm$, its rank equals the sum of the $m_j$s.

By prescribing specific spectral data, we can define various spaces of density operators acting on an $n$-dimensional Hilbert space $\HH$. Let $\bfp$ be an eigenvalue spectrum with associated degeneracy spectrum $\bfm$. We denote by $\DD(\bfp, \bfm)$ the space of all density operators on $\HH$ that have eigenvalues $\bfp$ with corresponding multiplicities given by $\bfm$. The space $\DD(\bfp, \bfm)$ is the unitary orbit of each of its elements: for any $\rho$ in $\DD(\bfp, \bfm)$ we have that
\begin{equation}
    \DD(\bfp,\bfm) = \{ U\rho U^\dagger : U \in \UU(\HH) \}.
\end{equation}
We say that two density operators are isospectral if and only if they belong to the same space $\DD(\bfp,\bfm)$, that is, if they have identical eigenvalue spectra including multiplicities.

The space $\DD(\bfPi,\bfm)$ consists of all density operators on $\HH$ with a common eigenprojector spectrum $\bfPi$. While the eigenvalues of these operators may vary, they all exhibit the same multiplicity structure given by $\bfm$. Moreover, they all commute. We assume the degeneracy spectrum $\bfm$ is compatible with $\bfPi$, meaning that the rank of the $j$th projector in $\bfPi$ equals the $j$th multiplicity $m_j$ in $\bfm$. The space $\DD(\bfPi, \bfm)$ is a smooth manifold canonically diffeomorphic to the open simplex~
\begin{equation}
    \big\{\boldsymbol{x} \in \mathbb{R}^l : x_j > x_{j+1} >0, \; \sum_{j=1}^l x_j m_j=1\big\},
\end{equation}
where $l$ is the common length of $\bfPi$ and $\bfm$. 

Given a degeneracy spectrum $\bfm$, we denote by $\DD(\bfm)$ the space of all density operators on $\HH$ whose (ordered) nonzero eigenvalues have multiplicities specified by $\bfm$. Two density operators in $\DD(\bfm)$ need not be isospectral nor share the same eigenprojector spectrum; however, they necessarily have the same rank. The space $\DD(\bfm)$ is a smooth manifold, with $\DD(\bfp,\bfm)$ and $\DD(\bfPi,\bfm)$ forming submanifolds~\cite{Gr2005, Gr2006}. Let $l$ denote the common length of $\bfp$, $\bfPi$, and $\bfm$, and let $r$ be the rank. The dimensions of these manifolds are
\begin{subequations}
\begin{align}
    \dim \DD(\bfp,\bfm) &= 2nr-r^2-\sum_{j=1}^l m_j^2, \\
    \dim \DD(\bfPi,\bfm) &= l-1, \\
    \dim \DD(\bfm) &= 2nr-r^2-\sum_{j=1}^l m_j^2+l-1. \label{dim of D(m)}
\end{align}
\end{subequations}

\subsection{Gauge theory for isodegenerate states}%
\label{sec: Gauge theory for isodegenerate states}
\noindent
Fix a degeneracy spectrum $\bfm = (m_1, m_2, \dots, m_l)$. We construct a principal bundle over $\DD(\bfm)$. For background on the geometric framework underlying this construction we refer the reader to Chaps.~9 and~10 of Ref.~\cite{Na2003}.

Let $r$ be the common rank of the density operators in $\DD(\bfm)$, and let $\KK$ be a Hilbert space of dimension $r$, henceforth referred to as the auxiliary Hilbert space. Fix an eigenprojector spectrum $\bfL = (\Lambda_1, \Lambda_2, \dots, \Lambda_l)$ of density operators on $\KK$ that is compatible with $\bfm$, meaning each $\Lambda_j$ is an orthogonal projection of rank $m_j$. We define $\WW(\bfL)$ as the space of linear maps $W:\KK\to\HH$ such that $W^\dagger W$ is a density operator on $\KK$ with eigenprojector spectrum $\bfL$.

\begin{figure}[t]
\centering
\includegraphics[width=0.9\linewidth]{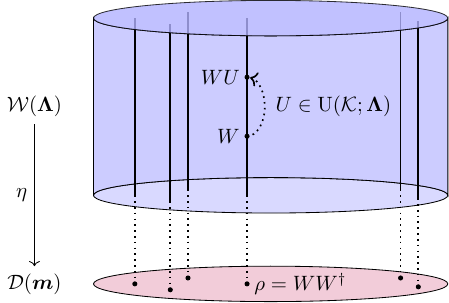}
\caption{The total space $\WW(\bfL)$ of the principal bundle $\eta$ over the space $\DD(\bfm)$ of density operators on $\HH$ with degeneracy spectrum $\bfm$ consists of all linear maps $W : \KK \to \HH$ such that $W^\dagger W$ is a density operator on $\KK$ with eigenprojectors given by $\bfL$. The complementary product $WW^\dagger$ belongs to $\DD(\bfm)$, and the bundle projection $\eta$ maps $W$ to $WW^\dagger$. The gauge group $\UU(\KK;\bfL)$, acting on $\WW(\bfL)$ from the right via operator composition, consists of the unitary operators on $\KK$ that commute with every projector in $\bfL$. This action is free and transitive on the fibers of the bundle, so each fiber is diffeomorphic to $\UU(\KK;\bfL)$.}
\label{fig: noll}
\end{figure}

As shown in Ref.~\cite{AnHe2015} (see also Ref.~\cite{An2018}), the map
\begin{equation}
    \eta : \WW(\bfL) \to \DD(\bfm), \quad W\mapsto WW^\dagger,
\end{equation}
is a principal fiber bundle. The gauge group consists of the unitary operators on $\KK$ that commute with all the
projection operators in~$\bfL$:
\begin{equation}
    \UU(\KK;\bfL) = \{ U\in \UU(\KK) : [U,\Lambda_j]=0,\, j=1,2\dots,l \}.
\end{equation}
This group acts on $\mathcal{W}(\bfL)$ from the right through operator composition:
\begin{equation}
    \WW(\bfL) \times \UU(\KK;\bfL) \to \WW(\bfL), \quad (W,U) \mapsto WU.
\end{equation}
We follow Uhlmann \cite{Uh1986, Uh1991, Uh1995} in referring to the elements of $\WW(\bfL)$ as amplitudes.\footnote{In Refs.~\cite{AnHe2015, An2018}, it is shown that $\eta$ arises naturally via symplectic reduction of the Uhlmann purification bundle.} Figure \ref{fig: noll} provides a schematic representation of the bundle $\eta$.

The collection of kernels of the differential of $\eta$ defines a subbundle of the tangent bundle of $\WW(\bfL)$, called the vertical bundle. The vertical space at an amplitude $W$ consists of all tangent vectors $\dot W$ at $W$ annihilated by the differential of $\eta$. This space can be expressed as
\begin{equation}
    \V_W\WW(\bfL) = \{ WX : X\in \uu(\KK;\bfL) \},
\end{equation}
where $\uu(\KK;\bfL)$ is the Lie algebra of $\UU(\KK;\bfL)$. This Lie algebra comprises the skew-Hermitian operators on $\KK$ that commute with every projection operator in $\bfL$.

\begin{figure}[t]
\centering
\includegraphics[width=0.95\linewidth]{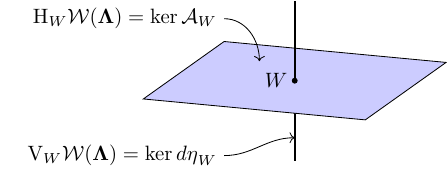}
\caption{The vertical space at $W$ consists of the tangent vectors at $W$ annihilated by $d\eta$ and is shown here as a vertical line. Its dimension is $m_1^2 + m_2^2 + \cdots + m_l^2$, equal to that of the fiber of $\eta$ through $W$, which is diffeomorphic to the gauge group. The horizontal space at $W$, orthogonal to the vertical space, has the same dimension as $\DD(\bfm)$; see Eq.~\eqref{dim of D(m)}. It is defined as the kernel of the connection form $\AA$ at $W$. Thus, a tangent vector $\dot{W}$ at $W$ is horizontal if $\AA(\dot W)=0$.}
\label{fig: ett}
\end{figure}

We equip $\WW(\bfL)$ with the Riemannian metric
\begin{equation}
    G(\dot W_1, \dot W_2) 
    = \frac{1}{2} \tr\big(\dot W_1^\dagger \dot W_2 + \dot W_2^\dagger \dot W_1\big),
\end{equation}
and define the horizontal bundle to be the orthogonal complement of the vertical bundle with respect to this metric. The horizontal space at an amplitude $W$ is
\begin{multline}
    \H_W\WW(\bfL) 
    = \{ \dot W \in \T_W \WW(\bfL) :
    \\ G(\dot W,WX)=0 \text{ for all } X\in\uu(\KK;\bfL) \}.
\end{multline}
Since the metric $G$ is gauge invariant, that is, the gauge group acts through isometries, this choice of horizontal distribution defines a connection on $\eta$. The corresponding $\mathfrak{u}(\KK;\bfL)$-valued connection form is given by
\begin{equation}
    \AA(\dot W) 
    = \frac{1}{2} \sum_{j=1}^l \Lambda_j \big( W^+\dot W - (W^+\dot W)^\dagger \big) \Lambda_j.
    \label{eq: Ehresmann connection}
\end{equation}
In this expression, $\dot{W}$ denotes an arbitrary tangent vector at the amplitude $W$, and $W^+$ is the Moore--Penrose pseudoinverse of $W$ \cite{Mo1920, Pe1955}:
\begin{equation}
    W^+ = (W^\dagger W)^{-1} W^\dagger.
    \label{eq: Moore--Penrose inverse}
\end{equation}
The horizontal bundle is the kernel bundle of the connection form, meaning that each horizontal subspace consists of the tangent vectors annihilated by the connection. Thus, the horizontal subspace at $W$ can equivalently be characterized as $\H_W\WW(\bfL) = \{ \dot W \in \T_W \WW(\bfL) : \AA(\dot W)=0 \}$. Figure~\ref{fig: ett} illustrates the decomposition of the tangent space at $W$ into horizontal and vertical subspaces.

\begin{ex}
\label{rmk: Hopf bundle}
If $\bfm = (1)$, the space $\DD(\bfm)$ coincides with the projective Hilbert space consisting of all rank-one orthogonal projection operators on $\HH$. These operators represent the pure states of the system. In this case, the amplitude space can be identified with the unit sphere in $\HH$, and the bundle $\eta$ equipped with the connection $\AA$ is isomorphic to the Hopf bundle endowed with the Aharonov--Anandan connection~\cite{AhAn1987}. Within this setting, a tangent vector $\ket{\dot\psi}$ of the sphere at $\ket{\psi}$ is horizontal if and only if $\AA(\ket{\dot\psi})=\braket{\psi}{\dot\psi}=0$. This condition is precisely the Aharonov--Anandan horizontality condition~\eqref{eq: parallel}.
\end{ex}

A detailed derivation of the connection form \eqref{eq: Ehresmann connection} is given in Appendix~\ref{app: The connection form}. In Appendix~\ref{app: Independence of the eigenprojector spectrum}, we show that the isomorphism class of $(\eta,\AA)$ is independent of both the choice of auxiliary Hilbert space $\KK$ and the eigenprojector spectrum $\bfL$. Consequently, for a given density operator $\rho$, we may take $\KK$ to be the support of $\rho$ and choose $\bfL$ to be its eigenprojector spectrum.

\subsection{Holonomy}
\label{sec: Holonomy}
\noindent
Let $\rho_t$ be a curve in $\DD(\bfm)$ starting at $\rho$. A standard result from the theory of principal fiber bundles \cite{Na2003} ensures that for any amplitude $W$ in the fiber over $\rho$, there exists a unique curve of amplitudes $W_t$ that starts at $W$, projects onto $\rho_t$, and has a horizontal velocity field:
    $\AA(\dot{W}_t) = 0$.
This curve is the \emph{horizontal lift} of $\rho_t$ that starts at $W$.

\begin{figure}[t]
\centering
\includegraphics[width=0.85\linewidth]{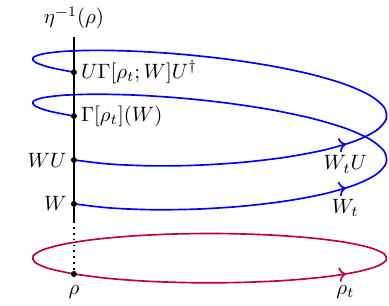}
\caption{Let $\rho_t$ be a closed curve in $\DD(\bfm)$ based at $\rho$, and let $W$ be an amplitude of $\rho$. The horizontal lift $W_t$ of $\rho_t$ starting at $W$ ends in the fiber above $\rho$ but need not return to $W$. This phenomenon is known as holonomy. The parallel transport operator maps $W$ to the endpoint of the horizontal lift: $\Gamma[\rho_t](W) = W_\tau$. If $U$ belongs to the gauge group, the horizontal lift starting at $WU$ is given by $W_t U$, so right multiplication by $U$ transfers the horizontal lift from $W$ to $WU$. Consequently, the parallel transport operator satisfies $\Gamma[\rho_t](WU) = \Gamma[\rho_t](W) U$, meaning it commutes with the gauge group action. The parallel transport of $WU$ can also be written as $U \Gamma[\rho_t;W] U^\dagger$, where $\Gamma[\rho_t;W]$ is the unique gauge group element transforming $W$ to $\Gamma[\rho_t](W)$.}
\label{fig: tva}
\end{figure}

Since the horizontal lift of $\rho_t$ is unique once the initial amplitude is specified, we can define a map from the fiber over the initial density operator $\rho$ to the fiber over the final density operator $\rho_\tau$, 
\begin{equation}
    \Gamma[\rho_t]:\eta^{-1}(\rho) \to \eta^{-1}(\rho_\tau), \quad W \mapsto \Gamma[\rho_t](W),
\end{equation}
by defining $\Gamma[\rho_t](W)$ to be the final amplitude of the unique horizontal lift of $\rho_t$ that starts at $W$.
This map is the \emph{parallel transport operator} associated with $\rho_t$.
If the curve $\rho_t$ is closed, so that $\rho_\tau = \rho$, the parallel transport operator maps the fiber over $\rho$ back to itself; see Figure~\ref{fig: tva}. In this case, the parallel transport operator is referred to as the \emph{holonomy} of $\rho_t$. 

\subsubsection{The holonomy group}
\begin{figure}[t]
\centering
\includegraphics[width=0.95\linewidth]{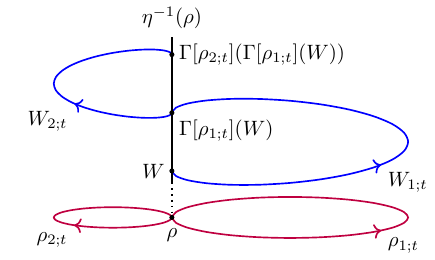}
\caption{The concatenation of two closed curves, $\rho_{1;t}$ and $\rho_{2;t}$, at $\rho$ defines a new closed curve $\rho_{1;t} \ast \rho_{2;t}$ at $\rho$. Its horizontal lift to the amplitude $W$ is obtained by concatenating the horizontal lift $W_{1;t}$ of $\rho_{1;t}$ starting at $W$ with the horizontal lift $W_{2;t}$ of $\rho_{2;t}$ starting at $\Gamma[\rho_{1;t}](W)$. The resulting horizontal lift terminates at $\Gamma[\rho_{2;t}](\Gamma[\rho_{1;t}](W))$, which establishes the composition law~\eqref{eq: composition law} for the parallel transport operator.}
\label{fig: tre}
\end{figure}

\noindent
The collection of all holonomies arising from closed curves based at $\rho$ forms a group under composition, called the holonomy group at $\rho$. The composition of the holonomies associated with two closed curves $\rho_{1;t}$ and $\rho_{2;t}$ at $\rho$ equals the holonomy of their concatenation. If $\rho_{1;t}$ and $\rho_{2;t}$ are defined on the intervals $[0,\tau_1]$ and $[0,\tau_2]$, respectively, their concatenation is the curve
\begin{equation}
    \rho_{1;t}\ast\rho_{2;t}
    = \begin{cases} 
        \rho_{1;t} &\text{ for }\quad 0\leq t\leq \tau_1, \\ 
        \rho_{2;t-\tau_1} &\text{ for }\quad \tau_1\leq t\leq \tau_1+\tau_2,
      \end{cases}
\end{equation}
which is again a closed curve at $\rho$, and
\begin{equation}
    \Gamma[\rho_{2;t}]\Gamma[\rho_{1;t}]=\Gamma[\rho_{1;t}\ast\rho_{2;t}]. 
    \label{eq: composition law}
\end{equation}
See Figure~\ref{fig: tre} for an illustration of the group operation. The identity element of the group is the holonomy of the constant curve at $\rho$, and the inverse of the holonomy associated with $\rho_t$ is given by the holonomy of the time-reversed curve.

The holonomy group at $\rho$ can be identified, though not canonically, with a subgroup of the gauge group $\UU(\KK;\bfL)$. For each amplitude $W$ in the fiber over $\rho$, we define the holonomy of $\rho_t$ \emph{at} $W$ to be the unitary operator $\Gamma[\rho_t; W]$ in the gauge group determined by
\begin{equation}
    \Gamma[\rho_t](W)=W\Gamma[\rho_t;W].
    \label{eq: local holonomy}
\end{equation}
The assignment $\Gamma[\rho_t] \mapsto \Gamma[\rho_t; W]$ is a group monomorphism that identifies the holonomy group at $\rho$ with a subgroup of $\UU(\KK;\bfL)$. The amplitude-based holonomies transform covariantly under changes of amplitude (see Figure~\ref{fig: tva}):
\begin{equation}
    \Gamma[\rho_t;WU] = U\Gamma[\rho_t;W]U^\dagger, \quad U\in \UU(\KK;\bfL).
\end{equation}
Thus, they constitute conjugate subgroups of the gauge group. Multiplying both sides of Eq.~\eqref{eq: local holonomy} on the left by $W^+$ yields $\Gamma[\rho_t;W] = W^+\Gamma[\rho_t](W)$.

\begin{ex}
    For pure states, $\eta$ reduces to the Hopf bundle over the projective Hilbert space (see Example~\ref{rmk: Hopf bundle}). In this case, the fiber over $\rho$ consists of all unit vectors $\ket{\psi}$ in the support of $\rho$, that is, those satisfying $\rho = \ketbra{\psi}{\psi}$. If $\rho_t$ is a curve of pure states starting and ending at $\rho$, the associated holonomy is given by $\Gamma[\rho_t](\ket{\psi}) = e^{i\theta}\ket{\psi}$, where $e^{i\theta}$ is the Aharonov--Anandan geometric phase factor appearing on the right-hand side of Eq.~\eqref{eq: Aharonov--Anandan holonomy}. This factor depends on the curve $\rho_t$ but not on the choice of lift $\ket{\psi_t}$, and we may therefore identify the holonomy $\Gamma[\rho_t]$ with it, as in Eq.~\eqref{eq: Aharonov--Anandan holonomy}.
    \label{rmk: AA holonomy}
\end{ex}

\begin{ex}
Suppose $\bfm = (1,1,\dots,1)$, so that the states under consideration have nondegenerate spectra. During a cyclic evolution of a state $\rho$ in $\DD(\bfm)$, the pure states appearing in a spectral decomposition of $\rho$ themselves undergo cyclic evolution and accumulate geometric phase factors, as described in Example~\ref{rmk: AA holonomy}.
We take $\KK$ to be the support of $\rho$ (cf.~the paragraph following Example~\ref{rmk: Hopf bundle}), and let $\ket{\psi_j}$ be an eigenvector of $\rho$ corresponding to its $j$th nonzero eigenvalue $p_j$. The holonomy of the cyclic trajectory $\rho_t$ of $\rho$ at the amplitude $W=\sum_{j=1}^l \sqrt{p_j} \ketbra{\psi_j}{\psi_j}$ is then given by
\begin{equation}
	\Gamma[\rho_t;W]
	= \sum_{j=1}^l e^{i\theta_j}\ketbra{\psi_j}{\psi_j},
\end{equation}
where $l$ is the length of $\bfm$ and $e^{i\theta_j}$ is the geometric phase factor acquired along the closed trajectory of $\ketbra{\psi_j}{\psi_j}$.
\label{rmk: nondeg spectrum}
\end{ex}

\subsubsection{Holonomy invariants}
\label{sec: Holonomy invariants}
\noindent
Two closed curves based at the same density operator are said to be holonomy equivalent if their holonomies coincide. Any property shared by all holonomy-equivalent curves is called a holonomy invariant. Because the amplitude-based holonomies transform covariantly under changes of amplitude, every unitarily invariant property of an amplitude-based holonomy is a holonomy invariant. Let $f$ be any function depending only on the eigenvalues of unitary operators in $\UU(\KK;\bfL)$. Such functions are unitarily invariant; hence $\rho_t\mapsto f(\Gamma[\rho_t;W])$ is a holonomy invariant. Important examples include the Wilson loop, defined as the trace of the holonomy, and the geometric phase, discussed in the next section. In Section~\ref{sec: Isoholonomic inequalities} we introduce another holonomy invariant, which we call the isoholonomic bound.

\begin{rmk}
The holonomy of a closed curve is invariant under orientation-preserving reparameterizations. Therefore, when calculating the holonomy or a holonomy invariant of a curve, one may freely apply orientation-preserving reparameterizations to the curve.
\label{rmk: reverse}
\end{rmk}

\subsubsection{Geometric phase}%
\label{sec: Geometric phase}
\noindent
Suppose $\rho_t$ is a (not necessarily closed) curve in $\DD(\bfm)$ starting at a density operator $\rho$. We define the geometric phase of $\rho_t$ as
\begin{equation}
    \thetageo[\rho_t] = \arg \tr \big(W^\dagger \Gamma[\rho_t](W)\big),
\end{equation}
where $W$ is any amplitude of $\rho$. Similar to the Aharonov--Anandan geometric phase, $\theta_{\mathrm{geo}}[\rho_t]$ is defined modulo $2\pi$. Next, we demonstrate that this definition coincides with the geometric phase introduced by Sj\"oqvist~\emph{et al.}~\cite{SjPaEkAnErOiVe2000} and extended by Tong~\emph{et al.}~\cite{ToSjKwOh2004}.

The orthogonal projection operators in $\bfL$ have mutually orthogonal supports whose direct sum exhausts the auxiliary Hilbert space: $\KK = \supp \Lambda_1 \oplus \supp \Lambda_2 \oplus \cdots \oplus \supp \Lambda_l$.
We call an orthonormal basis of $\KK$ of the form
\begin{equation}
    \{ \ket{ja} : j=1,2,\dots,l;\ a=1,2,\dots,m_j\}
    \label{eq: adapted basis}
\end{equation}
with $\ket{j1}, \ket{j2}, \dots, \ket{jm_j}$ spanning the support of $\Lambda_j$, a $\bfL$-adapted basis. In Appendix \ref{app: Horizontal lifts and parallel transported spectral decompositions}, we show that any $\bfL$-adapted basis establishes a one-to-one correspondence between parallel-transported spectral decompositions and horizontal lifts of $\rho_t$. This correspondence is given by the relation
\begin{equation}
    W_t\ket{ja} = \sqrt{p_{j;t}}\, \ket{\psi_{ja;t}},
    \label{eq: one-to-one correspondence}
\end{equation}
where the $p_{j;t}$s are the positive eigenvalues of $\rho_t$. Thus, given a parallel-transported spectral decomposition as in Eq.~\eqref{eq: spectral decomposition}, one can define a horizontal lift $W_t$ by specifying that Eq.~\eqref{eq: one-to-one correspondence} describes the action of $W_t$ on the $\bfL$-adapted basis vectors:
\begin{equation}
    W_t = \sum_{j=1}^l\sum_{a=1}^{m_j} \sqrt{p_{j;t}}\, \ketbra{\psi_{ja;t}}{ja}.
    \label{eq: expansion of horizontal lift}
\end{equation}
Conversely, if $W_t$ is a horizontal lift of $\rho_t$, then the vectors 
\begin{equation}
    \ket{\psi_{ja;t}} = p_{j;t}^{-1/2}W_t\ket{ja}
    \label{eq: the parallel vectors}
\end{equation}
form a parallel-transported spectral decomposition of $\rho_t$, as described in Eq.~\eqref{eq: spectral decomposition}.

For a horizontal lift $W_t$ of $\rho_t$, with a parallel-transported spectral decomposition established through the relation in Eq.~\eqref{eq: one-to-one correspondence}, the geometric phase of $\rho_t$ is given by
\begin{equation}
\begin{split}
    &\thetageo[\rho_t]
	= \arg\tr( W_0^\dagger W_\tau) \\
    &=  \arg\tr\sum_{j=1}^l \sum_{k=1}^l\sum_{a=1}^{m_j} \sum_{b=1}^{m_k} \sqrt{p_{j;0}p_{k;\tau}}\, \ket{ja} \braket{\psi_{ja;0}}{\psi_{kb;\tau}} \bra{kb} \\
    &= \arg\sum_{j=1}^l \sum_{a=1}^{m_j} \sqrt{p_{j;0}p_{j;\tau}} \braket{\psi_{ja;0}}{\psi_{ja;\tau}}.
\end{split}
\end{equation}
This expression coincides with the expression for the geometric phase in Eq.~\eqref{eq: clgp}.

\subsection{Gauge theory for isospectral and spectrally bounded states}%
\label{sec: Gauge theory for isospectral states}
\noindent
For any amplitude $W$ in $\WW(\bfL)$, the density operators $W^\dagger W$ and $WW^\dagger$ share the same set of positive eigenvalues, including multiplicities. Consequently, if $\bfp$ is an eigenvalue spectrum compatible with $\bfm$, the amplitudes $W$ that project into the space of isospectral states $\DD(\bfp,\bfm)$ must satisfy the condition
\begin{equation}
    W^\dagger W = \sum_{j=1}^l p_j \Lambda_j.
\end{equation}
We denote the space of such amplitudes by $\WW(\bfp,\bfL)$. It is shown in Ref.~\cite{AnHe2015} (see also Ref.~\cite{An2018}) that $\eta$ restricts to a principal bundle over $\DD(\bfp, \bfm)$ with total space $\WW(\bfp,\bfL)$, gauge group $\UU(\KK; \bfL)$, and connection form $\AA$. In this case, one can express the connection form as
\begin{equation}
    \AA(\dot W) 
    = \sum_{j=1}^l p_j^{-1}\Lambda_j W^\dagger \dot W \Lambda_j, \quad \dot W \in \T_W \WW(\bfL).
    \label{eq: connection in isospectral case}
\end{equation}
See Appendix \ref{app: The connection form} for details.

Let $\bfa = (\alpha_1,\alpha_2,\dots,\alpha_l)$ be a sequence of nonnegative real numbers. A density operator in $\DD(\bfm)$ is said to be spectrally bounded by $\bfa$ if its eigenvalues satisfy
\begin{equation}
    p_j \geq \alpha_j, \quad j=1,2,\dots,l.
    \label{eq: boundary condition}
\end{equation}
We denote by $\DD(\bfa,\bfm)$ the subset of $\DD(\bfm)$ consisting of density operators that are spectrally bounded by $\bfa$. The motivation for introducing this set will become clear in the next section.

\begin{rmk}
Let $\rho$ be a density operator in $\DD(\bfm)$ that is spectrally bounded by $\bfa$. The holonomies associated with closed curves in $\DD(\bfa,\bfm)$ based at $\rho$ form a subgroup of the holonomy group at $\rho$.
\label{rmk: restricted}
\end{rmk}

\begin{rmk}
The amplitudes $W$ projecting into $\DD(\bfa,\bfm)$ are precisely those for which $W^\dagger W$ is a density operator on $\KK$ whose eigenvalue spectrum satisfies condition \eqref{eq: boundary condition}. We denote the space of such amplitudes by $\WW(\bfa,\bfL)$. Because condition \eqref{eq: boundary condition} is imposed with ``$\geq$'' rather than ``$>$'', both $\DD(\bfa,\bfm)$ and $\WW(\bfa,\bfL)$ are manifolds with corners. The bundle map $\eta$ restricts to a principal bundle over $\DD(\bfa,\bfm)$ with total space $\WW(\bfa,\bfL)$, gauge group $\UU(\KK;\bfL)$, and connection form $\AA$ as defined in Eq.~\eqref{eq: Ehresmann connection}.
\label{rmk: corners}
\end{rmk}

\begin{rmk}
The bundle $\eta$ also restricts to a principal bundle over  $\DD(\bfPi,\bfm)$. Since the eigenvectors remain constant along any curve in this space, every horizontal lift of a closed curve is itself closed. Consequently, all closed curves in  $\DD(\bfPi,\bfm)$ have trivial holonomy.
\label{rmk: trivial}
\end{rmk}

\section{Isoholonomic inequalities}%
\label{sec: Isoholonomic inequalities}
\noindent
The metric $G$ induces a Riemannian metric $g$ on $\DD(\bfm)$ which makes $\eta$ a Riemannian submersion. The metric $g$ is defined as follows: Suppose $\dot\rho_1$ and $\dot\rho_2$ are tangent vectors to $\DD(\bfm)$ at $\rho$. Choose any amplitude $W$ of $\rho$, and let $\dot W_1$ and $\dot W_2$ be the horizontal lifts of $\dot\rho_1$ and $\dot\rho_2$ to $W$. Then,
\begin{equation}
g(\dot\rho_1, \dot\rho_2) = G(\dot W_1, \dot W_2).
\end{equation}
The horizontal lifts $\dot W_1$ and $\dot W_2$ exist and are unique because $d\eta$ restricts to an isomorphism from the horizontal space at $W$ to the tangent space at $\rho$. Moreover, the value of the inner product on the right-hand side is independent of the choice of amplitude $W$ since the gauge group acts through isometries; see Figure \ref{fig: fyra}. In Appendix \ref{app: Independence of the eigenprojector spectrum}, we prove that the metric $g$ does not depend on the choice of eigenprojector spectrum $\bfL$.

\begin{figure}[t]
\centering
\includegraphics[width=0.9\linewidth]{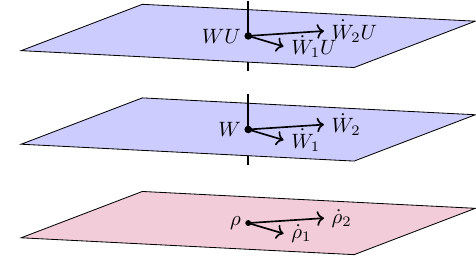}
\caption{The metric on $\DD(\bfm)$ is defined by lifting tangent vectors $\dot\rho_1$ and $\dot\rho_2$ at $\rho$ to horizontal vectors $\dot W_1$ and $\dot W_2$ at any amplitude $W$ of $\rho$, and evaluating their inner product $G(\dot W_1, \dot W_2)$. Because the gauge group acts by isometries, the result is independent of the choice of horizontal space used for the lift: the differential of the right action by any $U$ in $\UU(\KK;\bfL)$ maps $\dot W_1$ and $\dot W_2$ to $\dot W_1 U$ and $\dot W_2 U$, respectively, which still project to $\dot\rho_1$ and $\dot\rho_2$, and $G(\dot W_1 U, \dot W_2 U) = G(\dot W_1, \dot W_2)$.}
\label{fig: fyra}
\end{figure}

\begin{ex}
For pure states the induced metric coincides with the Fubini--Study metric: $g(\dot\rho_1,\dot\rho_2)=\frac{1}{2}\tr(\dot\rho_1\dot\rho_2)$.
\label{ex: FS}
\end{ex}

\subsection{Length and kinetic energy}%
\label{sec: Length and kinetic energy}
\noindent
An important property of Riemannian submersions is that they preserve the lengths and kinetic energies of horizontal curves. Consequently, for any curve $\rho_t$ in $\DD(\bfm)$ and any horizontal lift $W_t$ of $\rho_t$, the corresponding lengths and kinetic energies coincide: $\length[\rho_t] = \length[W_t]$ and $\energy[\rho_t] = \energy[W_t]$. These quantities are defined as follows:
\begin{subequations}
\begin{align}
    \length[\rho_t]
    &= \int_0^\tau \sqrt{g(\dot\rho_t,\dot\rho_t)}\,\dt, \\
    \energy[\rho_t]
    &= \frac{1}{2}\int_0^\tau g(\dot\rho_t,\dot\rho_t)\,\dt, \\
    \length[W_t]
    &= \int_0^\tau \sqrt{G(\dot W_t,\dot  W_t)}\,\dt, \\
    \energy[W_t]
    &= \frac{1}{2}\int_0^\tau G(\dot W_t,\dot  W_t)\,\dt.
\end{align}
\end{subequations}
The Cauchy--Schwarz inequality implies
\begin{equation}
    \length[\rho_t]^2 \leq 2\tau\energy[\rho_t].
    \label{eq: CS}
\end{equation}
While the length of $\rho_t$ is invariant under reparameterization, the kinetic energy is not. Equality in Eq.~\eqref{eq: CS} holds if and only if $\rho_t$ is parameterized proportionally to arc length, that is, if it has constant speed.

\subsubsection{Fisher--Rao length and kinetic energy}
\label{sec: Fisher--Rao length and kinetic energy}
\noindent
Assume that the eigenvalues of $\rho_t$ changes over time, $\bfp_t=(p_{1;t},p_{2;t},\dots,p_{l;t})$. The time-varying vector
\begin{equation}
    \boldsymbol{P}_t=(m_1p_{1;t},m_2p_{2;t},\dots,m_lp_{l;t})
\end{equation}
defines a curve within the open simplex of positive probability distributions of length $l$. The Fisher--Rao metric induces a natural Riemannian structure on this simplex \cite{Fi1925, Ra1992}, and the squared Fisher--Rao speed of $\boldsymbol{P}_t$ is
\begin{equation}
    \gFR(\dot{\boldsymbol{P}}_t,\dot{\boldsymbol{P}}_t) = \frac{1}{4} \sum_{j=1}^l m_j \frac{\dot p_{j;t}^2}{p_{j;t}}.
\end{equation}
We write $\lengthFR[\bfp_t,\bfm]$ and $\energyFR[\bfp_t,\bfm]$ for its Fisher--Rao length and kinetic energy, respectively:
\begin{subequations}
\begin{align}
    \lengthFR[\bfp_t,\bfm] &= \frac{1}{2}\int_0^\tau \sqrt{\sum_{j=1}^l m_j \frac{\dot p_{j;t}^2}{p_{j;t}}}\,\dt, \\
    \energyFR[\bfp_t,\bfm] &= \frac{1}{8}\int_0^\tau \sum_{j=1}^l m_j \frac{\dot p_{j;t}^2}{p_{j;t}}\,\dt.
\end{align}
\end{subequations}
These quantities capture the statistical length and kinetic energy of the evolution of the eigenvalue spectrum of $\rho_t$.

\begin{rmk}
    When a curve of isodegenerate states is reparameterized, its curve of eigenvalues is also reparameterized. This, however, does not affect the Fisher--Rao length of the corresponding curve of probability distributions. 
    \label{rmk: FR}
\end{rmk}

\subsection{The isoholonomic inequality for isospectral dynamics}%
\label{sec: The isoholonomic inequality for isospectral dynamics}
\noindent
Consider the bundle $\eta$ over the space of density operators that share an eigenvalue spectrum $\bfp$.
Since the unitary operators in the gauge group commute with the operators in $\bfL$, any such unitary operator $U$ can be decomposed as
\begin{equation}
    U=U_1\oplus U_2\oplus\cdots \oplus U_l,
    \label{eq: decomposition of gauge group unitary}
\end{equation}
where $U_j$ is a unitary operator on the support of $\Lambda_j$.\footnote{Thus, $U$ has a block-diagonal matrix representation relative to any $\bfL$-adapted basis, with the $j$th block being an $m_j \times m_j$ submatrix.} We define a unitarily invariant function on the gauge group as follows: For a unitary $U$ with a decomposition as in Eq.~\eqref{eq: decomposition of gauge group unitary}, let $\theta_{j1},\theta_{j2},\dots,\theta_{jm_j}$ be the phases of the eigenvalues of $U_j$ within the interval $[0,2\pi)$. Then, define
\begin{equation}
    \isobound(\bfp;U) = \sqrt{ \sum_{j=1}^l \sum_{a=1}^{m_j} p_j \theta_{ja}(2\pi - \theta_{ja}) }.
\end{equation}
The function $U\mapsto \isobound(\bfp;U)$ is unitarily invariant because $\isobound(\bfp;U)$ depends only on the spectrum of $U$.

Fix a density operator $\rho$ in $\DD(\bfp,\bfm)$, and let $W$ be any amplitude of $\rho$. Since the holonomies at $W$ of closed curves in $\DD(\bfp,\bfm)$ at $\rho$ are elements in the gauge group, and $U\mapsto\isobound(\bfp;U)$ is a unitarily invariant function on this group,
\begin{equation}
    \isobound[\rho_t] = \isobound(\bfp;\Gamma[\rho_t;W])
\end{equation}
is a holonomy invariant. We call this invariant the isoholonomic bound. The isoholonomic inequality states that the length of a closed curve $\rho_t$ in $\DD(\bfp,\bfm)$ is bounded from below by its isoholonomic bound:
\begin{equation}
    \length[\rho_t] \geq \isobound[\rho_t].
    \label{eq: insoholonomic inequality for isospectral states}
\end{equation}
More generally, the lengths of all closed curves in $\DD(\bfp,\bfm)$ at $\rho$ with a given holonomy are bounded from below by the isoholonomic bound associated with that holonomy. In Section~\ref{sec: Tightness of the isoholonomic inequality for isospectral dynamics} we show that if the dimension of the Hilbert space is at least twice the common rank of the density operators in $\DD(\bfp,\bfm)$, then the isoholonomic inequality~\eqref{eq: insoholonomic inequality for isospectral states} is tight. Specifically, for every unitary in the gauge group there exists a closed curve at $\rho$ whose holonomy equals that unitary and whose length attains the corresponding isoholonomic bound.

\begin{rmk}
The notion of an isoholonomic inequality was introduced by Montgomery as a partial response to the isoholonomic problem formulated in Ref.~\cite{Mo1990}:
\begin{center}
    \textit{``Among all loops with a fixed holonomy, \\ find the one with shortest length.''}    
\end{center}
\label{rmk: Montgomery}
\end{rmk}

\begin{ex}
Consider a qubit initially in the state
\begin{equation}
\rho = p_0 \ketbra{0}{0} + p_1 \ketbra{1}{1},
\end{equation}
where $\ket{0}$ and $\ket{1}$ form an orthonormal basis of the qubit Hilbert space $\HH$, and $p_0$ and $p_1$ are distinct and nonzero. Suppose the state evolves according to $\rho_t = U_t \rho U_t^\dagger$, where
\begin{equation}
	U_t = \cos\Big(\frac{\omega t}{2}\Big)\1 - i\sin\Big(\frac{\omega t}{2}\Big)\boldsymbol{n}\cdot \boldsymbol{\sigma}.
	\label{eq: qubit time-evolution operator}
\end{equation}
Here $\boldsymbol{n} = (n_1,n_2,n_3)$ is a unit vector in $\mathbb{R}^3$, assumed not to be parallel to $(0,0,1)$ so that $\rho_t$ is nonstationary, and $\boldsymbol{\sigma} = (\sigma_1,\sigma_2,\sigma_3)$ is the triple of Pauli operators defined by
\begin{subequations}
\begin{align}
	\sigma_1 &= \ketbra{1}{0} + \ketbra{0}{1}, \\
	\sigma_2 &= i(\ketbra{1}{0} - \ketbra{0}{1}), \\
	\sigma_3 &= \ketbra{0}{0} - \ketbra{1}{1}.
\end{align}
\end{subequations}
In this case we may take $\KK=\HH$ and $\bfL=(\ketbra{0}{0},\ketbra{1}{1})$. The evolution is periodic with period $\tau = 2\pi/\omega$, and the holonomy associated with the curve traced out over one period is given by
\begin{equation}
	\Gamma[\rho_t] = e^{i\pi(1+n_3)}\ketbra{0}{0} + e^{i\pi(1-n_3)}\ketbra{1}{1}.
\end{equation}
The phases of the eigenvalues of the holonomy in the interval $[0,2\pi)$ are therefore $\theta_0 = \pi(1+n_3)$ and $\theta_1 = \pi(1-n_3)$, and the corresponding isoholonomic bound is
\begin{equation}
\begin{split}
	\isobound[\rho_t]
	= \sqrt{\sum_{j=0}^1 p_j \theta_j (2\pi - \theta_j)}
	= \pi \sqrt{1 - n_3^2}.
\end{split}
\label{eq: femtifem}
\end{equation}
In Example~\ref{ex: speed limit mixed qubit} we show that the length of $\rho_t$ attains the isoholonomic bound.
\label{ex: mixed qubit}
\end{ex}

\subsection{The isoholonomic inequality for spectrally constrained isodegenerate dynamics}%
\label{The isoholonomic inequality for spectrally constrained isodegenerate dynamics}
\noindent
If we allow for the eigenvalue spectrum to vary with time, we need to adjust the definition of the isoholonomic bound accordingly. In this case, we restrict our focus to density operators whose eigenvalue spectrum is constrained by a sequence $\bfa$, as described in Eq.~\eqref{eq: boundary condition}. For a unitary $U$ with a decomposition as described in Eq.~\eqref{eq: decomposition of gauge group unitary}, we define
\begin{equation}
    \isobound(\bfa;U)= \sqrt{\sum_{j=1}^l \sum_{a=1}^{m_j} \alpha_j \theta_{ja}(2\pi - \theta_{ja})},
\end{equation}
where, as before, $\theta_{j1},\theta_{j2},\dots,\theta_{jm_j}$ are the phases of the eigenvalues of $U_j$ within the interval $[0,2\pi)$. Now, let $W$ be any amplitude of $\rho$ in $\DD(\bfa,\bfm)$. Then
\begin{equation}
    \isobound[\bfa;\rho_t] = \isobound(\bfa;\Gamma[\rho_t;W])
\end{equation}
is a holonomy invariant. We call this invariant the $\bfa$-constrained isoholonomic bound of $\rho_t$.

The length of any closed curve $\rho_t$ in $\DD(\bfa,\bfm)$ is lower bounded by its $\bfa$-constrained isoholonomic bound:
\begin{equation}
    \length[\rho_t]\geq \isobound[\bfa;\rho_t].
    \label{eq: alpha-isoholonomic inequality}
\end{equation}
However, in this case, a stronger inequality can be established: Let $\bfp_t$ be the eigenvalue spectrum of $\rho_t$. Then,~
\begin{equation}
    \length[\rho_t]^2 \geq \lengthFR[\bfp_t,\bfm]^2 + \isobound[\bfa;\rho_t]^2.
    \label{eq: strong isoholonomic inequality}
\end{equation}
We conclude that the isoholonomic inequality \eqref{eq: alpha-isoholonomic inequality} is strict whenever the eigenvalue spectrum of $\rho_t$ varies with time, since in that case $\lengthFR[\bfp_t,\bfm]$ is positive. If, on the other hand, the eigenvalue spectrum of $\rho_t$ remains constant, then the length of $\rho_t$ is bounded below by $\isobound[\rho_t]$, which, in view of the bounds \eqref{eq: boundary condition}, is greater than or equal to $\isobound[\bfa;\rho_t]$.

\subsection{Derivations of the isoholonomic inequalities}%
\label{sec: Derivations of the isoholonomic inequalities}
\noindent
Let $\rho_t$ be a curve in $\DD(\bfm)$ that starts and ends at $\rho$. Reparameterize it proportionally to arc length while maintaining its orientation.\footnote{Such a reparameterization preserves both the length and the holonomy of $\rho_t$.}
Let $W$ be an amplitude of $\rho$. The holonomy of $\rho_t$ at $W$ admits the decomposition
\begin{equation}
    \Gamma[\rho_t;W] 
    = \Gamma[\rho_t;W]_1 \oplus \Gamma[\rho_t;W]_2 \oplus \cdots \oplus \Gamma[\rho_t;W]_l,
\end{equation}
where each $\Gamma[\rho_t;W]_j$ is a unitary operator acting on the support of $\Lambda_j$. Let $\theta_{j1},\theta_{j2},\dots,\theta_{jm_j}$ denote the phases of the eigenvalues of $\Gamma[\rho_t;W]_j$ in $[0,2\pi)$, and fix a $\bfL$-adapted basis as specified in Eq.~\eqref{eq: adapted basis} such that each $\ket{ja}$ is an eigenvector of $\Gamma[\rho_t;W]_j$, that is, $\Gamma[\rho_t;W]_j \ket{ja} = e^{i\theta_{ja}} \ket{ja}$. Let $W_t$ be the horizontal lift of $\rho_t$ starting at $W$, define $\ket{\psi_{ja;t}}$ as in Eq.~\eqref{eq: the parallel vectors}, and set $\rho_{ja;t} = \ketbra{\psi_{ja;t}}{\psi_{ja;t}}$. Because $\ket{ja}$ is an eigenvector of the holonomy, the final vector $\ket{\psi_{ja;\tau}}$ differs from the initial vector $\ket{\psi_{ja;0}}$ only by a phase factor:
\begin{equation}
\begin{split}
    \ket{\psi_{ja;\tau}}
    &= p_{j;\tau}^{-1/2} W_\tau\ket{ja} \\
    &= p_{j;\tau}^{-1/2}W\Gamma[\rho_t;W]\ket{ja} \\
    &= e^{i\theta_{ja}} p_{j;0}^{-1/2}W\ket{ja} \\
    &= e^{i\theta_{ja}}\ket{\psi_{ja;0}}. 
\end{split}   
\end{equation}
Hence each $\rho_{ja;t}$ is a closed curve. Since the vectors $\ket{\psi_{ja;t}}$ satisfy the parallel-transport condition~\eqref{eq: degenerate parallel transport condition}, they are horizontal, and $\theta_{ja}$ is the Aharonov--Anandan geometric phase associated with $\rho_{ja;t}$. The isoholonomic inequality for pure-state evolutions~\eqref{eq: isoholonomic inequality for pure states} therefore implies
\begin{equation}
    \lengthFS[\rho_{ja;t}]^2 \geq \theta_{ja}(2\pi - \theta_{ja}).
    \label{eq: sextiotre}
\end{equation}

Because $\rho_t$ is parameterized proportionally to arc length and the kinetic energies of $\rho_t$ and $W_t$ coincide, we have
\begin{equation}
    \length[\rho_t]^2
    = 2\tau \energy[\rho_t]
    = 2\tau \energy[W_t].
    \label{eq: sextiofyra}
\end{equation}
The kinetic energy of $W_t$ decomposes into two contributions, one of which is the Fisher--Rao kinetic energy of the eigenvalue distribution of $\rho_t$:
\begin{equation}
    \energy[W_t] 
    = \energyFR[\bfp_t,\bfm]
    + \frac{1}{2} \sum_{j=1}^l \sum_{a=1}^{m_j} \int_0^\tau \! p_{j;t} \braket{\dot\psi_{ja;t}}{\dot\psi_{ja;t}}\,\dt.
\label{eq: sextiofem}
\end{equation}

Now, suppose that $\rho_t$ is an evolution through isospectral density operators, so that $\bfp_t=\bfp$ for all $t$. Then the kinetic energy of the eigenvalue distribution is zero, and each term in the double sum in Eq.~\eqref{eq: sextiofem} can be rewritten as
\begin{equation}
    \frac{1}{2}  \int_0^\tau p_{j;t} \braket{\dot\psi_{ja;t}}{\dot\psi_{ja;t}}\,\dt 
    =  p_j\energyFS[\rho_{ja;t}].
    \label{eq: sextiosex}
\end{equation}
By combining the results from Eqs.~\eqref{eq: sextiotre}--\eqref{eq: sextiosex}, we obtain the following:
\begin{equation}
\begin{split}
    \length[\rho_t]^2
    &= \sum_{j=1}^l \sum_{a=1}^{m_j} 2\tau p_j\energyFS[\rho_{ja;t}] \\
    &\geq \sum_{j=1}^l \sum_{a=1}^{m_j} p_j \lengthFS[\rho_{ja;t}]^2 \\
    &\geq \sum_{j=1}^l \sum_{a=1}^{m_j} p_j \theta_{ja}(2\pi - \theta_{ja}) \\
    &= \isobound[\rho_t]^2.
\end{split}
\end{equation}
This proves the isoholonomic inequality \eqref{eq: insoholonomic inequality for isospectral states}.

If the eigenvalue spectrum of $\rho_t$ varies with time but is constrained by the sequence $\bfa$ as specified in Eq.~\eqref{eq: boundary condition}, we can bound the terms in the double sum in Eq.~\eqref{eq: sextiofem} from below as follows:
\begin{equation}
    \frac{1}{2} \int_0^\tau p_{j;t} \braket{\dot\psi_{ja;t}}{\dot\psi_{ja;t}}\dt 
    \geq \alpha_j \energyFS[\rho_{ja;t}].
\end{equation}
It follows that
\begin{equation}
\begin{split}
    \length[\rho_t]^2
    &\geq 2\tau\energyFR[\bfp_t,\bfm] 
      + \sum_{j=1}^l \sum_{a=1}^{m_j} 2\tau \alpha_j \energyFS[\rho_{ja;t}] \\
    &\geq \lengthFR[\bfp_t,\bfm]^2
      + \sum_{j=1}^l \sum_{a=1}^{m_j} \alpha_j \lengthFS[\rho_{ja;t}]^2 \\
    &\geq \lengthFR[\bfp_t,\bfm]^2
      + \sum_{j=1}^l \sum_{a=1}^{m_j} \alpha_j \theta_{ja}(2\pi - \theta_{ja}) \\
    &= \lengthFR[\bfp_t,\bfm]^2 + \isobound[\bfa;\rho_t]^2.
\end{split}
\end{equation}
This establishes Eq.~\eqref{eq: strong isoholonomic inequality} and thereby proves the isoholonomic inequality~\eqref{eq: alpha-isoholonomic inequality}.

\begin{rmk}
Reparameterizing $\rho_t$ affects the parameterization of the eigenvalue spectrum. Therefore, $\energyFR[\bfp_t,\bfm]$ does not necessarily correspond to the kinetic energy of the probability distribution associated with the original time-varying spectrum of $\rho_t$. However, the value of the Fisher--Rao length functional remains unchanged under reparameterization. This means that $\lengthFR[\bfp_t,\bfm]$ \emph{is} the Fisher--Rao length of the probability distribution associated with the original time-varying spectrum of $\rho_t$.
\end{rmk}

\section{Speed limit for the return time of unitarily evolving cyclic systems}%
\label{sec: Speed limit for the return time of unitarily evolving cyclic systems}
\noindent
Assume that the state evolves unitarily along a closed curve $\rho_t = U_t\rho U_t^\dagger$, where $U_t$ is the time propagator generated by a Hamiltonian $H_t$. Let $\bfp$ be the constant eigenvalue spectrum of the state, and write $\Delta E$ for the average energy uncertainty along the evolution:
\begin{equation}
    \Delta E = \frac{1}{\tau} \int_0^\tau \Delta H_t\, \dt, 
    \quad \Delta^2 H_t = \tr(\rho_t H_t^2)-\tr(\rho_t H_t)^2.
\end{equation}
In this section, we show that the total evolution time $\tau$ required for the system to return to its initial state is bounded from below by the isoholonomic bound associated with $\rho_t$, divided by the average energy uncertainty:
\begin{equation}
    \tau \geq \frac{\isobound[\rho_t]}{\Delta E}.
    \label{eq: MT bound}
\end{equation}

\begin{ex}
Consider the periodically evolving qubit discussed in Example~\ref{rmk: nondeg spectrum}. Assume that its evolution is generated by the Hamiltonian
\begin{equation}
	H = \frac{\omega}{2}\boldsymbol{n}\cdot\boldsymbol{\sigma},
\end{equation}
whose associated time-evolution operator is given in Eq.~\eqref{eq: qubit time-evolution operator}. The energy uncertainty of the system, which is conserved under the dynamics, is given by
\begin{equation}
	\Delta E = \frac{\omega}{2}\sqrt{1 - n_3^2 (p_0 - p_1)^2}.
\end{equation}
The corresponding speed limit for the period of this system is therefore
\begin{equation}
	\frac{\isobound[\rho_t]}{\Delta E}
	= \frac{2\pi}{\omega} \sqrt{\frac{1 - n_3^2}{1 - n_3^2 (p_0 - p_1)^2}}.
\end{equation}
We find that this speed limit coincides with the period $\tau = 2\pi/\omega$ if and only if $n_3 = 0$.
\label{ex: qubit Hamiltonian}
\end{ex}

\subsection{Derivation of the quantum speed limit}%
\label{sec: Derivation of the quantum speed limit}
\noindent
Let $W$ be any amplitude of $\rho$ and define a lift of $\rho_t$ as $W_t=U_tW$. This lift is not necessarily horizontal, since the velocity vector $\dot W_t = -iH_tW_t$ need not be orthogonal to the vertical space at $W_t$ for every $t$.

To derive the horizontal component of the velocity vector, let $\Pi_{j;t}$ be the orthogonal projection operator onto the eigenspace of $\rho_t$ corresponding to the eigenvalue $p_j$. Furthermore, if $\rho_t$ does not have full rank, let $\Pi_{0;t}$ be the orthogonal projection onto the orthogonal complement of the support of $\rho_t$; otherwise, let $\Pi_{0;t}$ be the zero operator. We then define the state-incoherent component and the state-coherent component of $H_t$ as
\begin{subequations}
\begin{align}
    H_t^{\inco} &= \sum_{j=0}^l \Pi_{j;t} H_t \Pi_{j;t}, \label{eq: incoherent part} \\
    H_t^{\co} &= H_t-H_t^\inco. \label{eq: coherent part}
\end{align}
\end{subequations}
Since the incoherent component commutes with $\rho_t$, the velocity vector field of $\rho_t$ reduces as
\begin{equation}
    \dot\rho_t = -i[H_t,\rho_t] = -i[H_t^\co, \rho_t].
\end{equation}
Moreover, the energy variance decomposes into two terms, corresponding to the variances of the coherent and incoherent components of the Hamiltonian:
\begin{equation}
\begin{split}
    \Delta^2H_t 
    &= \tr\big(\rho_t(H_t^\co)^2\big) + \tr\big(\rho_t\{H_t^\inco,H_t^\co\}\big) \\
    &\quad + \tr\big(\rho_t(H_t^\inco)^2\big) - \tr\big(\rho_t H_t^\co\big)^2 \\  
    &\quad- 2\tr\big(\rho_t H_t^\co\big)\tr\big(\rho_t H_t^\inco\big) - \tr\big(\rho_t H_t^\inco\big)^2 \\
    &= \tr\big(\rho_t(H_t^\co)^2\big)+\tr\big(\rho_t(H_t^\inco)^2\big)  - \tr\big(\rho_t H_t^\inco\big)^2 \\
    &= \Delta^2H_t^\co + \Delta^2H_t^\inco.
    \label{eq: uncertainty decomposition}
\end{split}
\end{equation}
For the second and third identities, we used that both $\tr(\rho_t\{H_t^\inco, H_t^\co\})$ and $\tr(\rho_t H_t^\co)$ are identically zero.

\begin{figure}[t]
\centering
\includegraphics[width=0.9\linewidth]{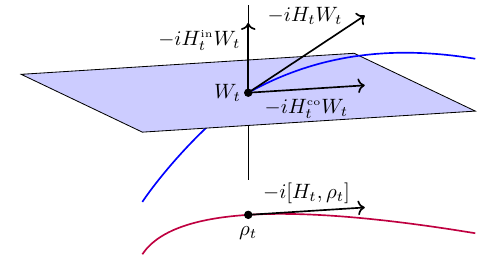}
\caption{The lift $W_t = U_t W$ of the curve $\rho_t = U_t \rho U_t^\dagger$ is not necessarily horizontal. The vertical component of the velocity vector $\dot{W}_t = -i H_t W_t$ is $-i H_t^\inco W_t$, and the horizontal component is $-i H_t^\co W_t$. The operators $H_t^\inco$ and $H_t^\co$ are the state-incoherent and state-coherent components of $H_t$. These are defined in Eqs.~\eqref{eq: incoherent part} and \eqref{eq: coherent part}, respectively. The speed of $\rho_t$ equals the uncertainty of the state-coherent part of $H_t$.}
\label{fig: fem}
\end{figure}

The variance of the coherent component equals the square of the speed of $\rho_t$. To see this, decompose the velocity vector field of $W_t$ as
\begin{equation}
    \dot W_t = -iH_t^\inco W_t-iH_t^\co W_t.
    \label{eq: decompostion}
\end{equation}
The first term on the right is vertical, as it lies in the kernel of the differential of the bundle projection:
\begin{equation}
    d\eta (-iH_t^\inco W_t) = -i[H_t^\inco,\rho_t]=0.
\end{equation}
To show that the second term is horizontal, expand $W_t$ as in Eq.~\eqref{eq: expansion of horizontal lift}, where the vectors $\ket{\psi_{ja;t}}$ form a parallel-transported spectral decomposition of $\rho_t$. Then,
\begin{equation}
\begin{split}
    \AA(-iH_t^\co W_t)
    &= -i \sum_{j=1}^l p_j^{-1}\Lambda_j W_t^\dagger H_t^\co W_t \Lambda_j \\ 
    &= -i \sum_{j=1}^l \sum_{a,b=1}^{m_j} \ketbra{ja}{\psi_{ja;t}} H_t^\co \ketbra{\psi_{jb;t}}{jb} \\
    &= 0.
\end{split}
\end{equation}
The last equality holds because $\Pi_{j;t} H_t^\co \Pi_{j;t} = 0$. Thus, the decomposition in Eq.~\eqref{eq: decompostion} is the decomposition of $\dot W_t$ into its vertical and horizontal components; see Figure \ref{fig: fem}.
It follows that the speed of $\rho_t$ equals the uncertainty of the state-coherent component of $H_t$:
\begin{equation}
\begin{split}
    g(\dot\rho_t,\dot\rho_t) 
    &= G(-iH_t^\co W_t,-iH_t^\co W_t) \\
    &= \tr\big(W_t^\dagger (H_t^\co)^2 W_t\big) \\
    &= \Delta^2H_t^\co.
    \label{eq: speed}
\end{split}
\end{equation}
Equations~\eqref{eq: uncertainty decomposition} and \eqref{eq: speed} imply that the energy uncertainty upper bounds the speed of $\rho_t$, yielding $\tau \Delta E \geq \length[\rho_t]$. The Mandelstam--Tamm type quantum speed limit~\eqref{eq: MT bound} then follows directly from the isoholonomic inequality~\eqref{eq: insoholonomic inequality for isospectral states}.

\begin{ex}
Consider the periodically evolving qubit discussed in Examples~\ref{rmk: nondeg spectrum} and~\ref{ex: qubit Hamiltonian}. The state-incoherent and state-coherent parts of the Hamiltonian are given by
\begin{subequations}
    \begin{align}
        H^\inco &= \frac{\omega}{2}\Big((1+n_3)\ketbra{0}{0} + (1-n_3)\ketbra{1}{1}\Big), \\
        H^\co &= \frac{\omega}{2}\Big((n_1+in_2)\ketbra{1}{0} + (n_1-in_2)\ketbra{0}{1}\Big).
    \end{align}
    \end{subequations}
Equation~\eqref{eq: speed} implies that the length of the trajectory over one period equals the period multiplied by the uncertainty of the state-coherent part of the Hamiltonian:
\begin{equation}
    \length[\rho_t] 
    = \tau \Delta H^\co
    = \frac{2\pi}{\omega} \sqrt{\tr(\rho(H^\co)^2)}
    = \pi\sqrt{1-n_3^2}.
\end{equation}
We conclude that the length of the state's evolution curve equals its isoholonomic bound; cf.~Eq.~\eqref{eq: femtifem}.
\label{ex: speed limit mixed qubit} 
\end{ex}

\section{Tightness of the isoholonomic inequality for isospectral dynamics}%
\label{sec: Tightness of the isoholonomic inequality for isospectral dynamics}
\noindent
Let $\bfp$ and $\bfm$ be compatible eigenvalue and degeneracy spectra, and assume that the sum of the degeneracies is at most half the dimension of the Hilbert space $\HH$. Fix a state $\rho$ with spectra $\bfp$ and $\bfm$, and choose an amplitude $W$ of $\rho$. Also, fix a final time $\tau > 0$.

In this section, we adapt an idea from Ref.~\cite{So2025b} to show that for any unitary operator $U$ in the gauge group, there exists a closed curve $\rho_t$ in $\DD(\bfp,\bfm)$ based at $\rho$ with holonomy $U$ at $W$, and whose length equals the isoholonomic bound $\isobound(\bfp;U)$. This curve thus saturates the isoholonomic inequality: $\length[\rho_t]=\isobound[\rho_t]$. More precisely, we show that there exists a Hamiltonian $H_t$ that unitarily generates the evolution $\rho_t$ from the initial state $\rho$, and that this Hamiltonian can be chosen to be state-coherent along the entire trajectory $\rho_t$. As a consequence, $\length[\rho_t] = \tau \Delta E$. This implies that the Mandelstam--Tamm type quantum speed limit~\eqref{eq: MT bound} is saturated.

\subsection{Proof of tightness}%
\label{sec: Proof of tightness}
\noindent
Let $U$ be an arbitrary unitary element of the gauge group. Decompose $U$ as in Eq.~\eqref{eq: decomposition of gauge group unitary}, and let $\theta_{j1}, \theta_{j2}, \dots, \theta_{jm_j}$ denote the phases of the eigenvalues of its $j$th component, chosen in the interval $[0,2\pi)$. Fix an $\bfL$-adapted basis of $\KK$ consisting of eigenvectors of $U$, so that
\begin{equation}
    U = \sum_{j=1}^l \sum_{a=1}^{m_j} e^{i\theta_{ja}} \ketbra{ja}{ja}.
    \label{eq: spectral of U}
\end{equation}
Define normalized, pairwise orthogonal vectors according to the correspondence in Eq.~\eqref{eq: one-to-one correspondence} by $\ket{\psi_{ja}} = p_j^{-1/2} W\ket{ja}$. For $j = 1, 2, \dots, l$ and $a = 1, 2, \dots, m_j$ choose mutually orthogonal two-dimensional subspaces $\HH_{ja}$ of $\HH$ such that the intersection of $\HH_{ja}$ with the support of $\rho$ is spanned by $\ket{\psi_{ja}}$.\footnote{This is where the assumption that the dimension of $\HH$ is at least twice the rank of $\rho$ is used.} Reference~\cite{So2025b} shows that there exists a Hamiltonian $H_t$ that vanishes on the orthogonal complement of the sum of the subspaces $\HH_{ja}$ and evolves each $\ket{\psi_{ja}}$ along a trajectory $\ket{\psi_{ja;t}}$ with the following properties:
\begin{enumerate}[label=\alph*),itemsep=2pt, topsep=3pt]
    \item The curve is Aharonov--Anandan horizontal.
    \item It evolves at constant speed $(2\pi\theta_{ja} - \theta_{ja}^2)^{1/2}/\tau$.
    \item It remains entirely within $\HH_{ja}$.
\end{enumerate}
Define $\rho_t$ and its corresponding lift $W_t$ as
\begin{subequations}
\begin{align}
    \rho_t&=\sum_{j=1}^l\sum_{a=1}^{m_j} p_j\ketbra{\psi_{ja;t}}{\psi_{ja;t}}, \\
    W_t&=\sum_{j=1}^l\sum_{a=1}^{m_j} \sqrt{p_j}\,\ketbra{\psi_{ja;t}}{ja}.
\end{align}
\end{subequations}
Then $\rho_t$ lies in $\DD(\bfp,\bfm)$ and starts and ends at $\rho$, while $W_t$ lies in $\WW(\bfp,\bfL)$ and starts at $W$. Since each trajectory $\ket{\psi_{ja;t}}$ is Aharonov--Anandan horizontal, and trajectories corresponding to distinct indices $a \neq b$ evolve in mutually orthogonal subspaces, the parallel-transport condition~\eqref{eq: degenerate parallel transport condition} is satisfied, ensuring that $W_t$ is horizontal.

To verify that $H_t$ is state-coherent, it suffices to show that its incoherent part vanishes. This follows from the parallel-transport condition~\eqref{eq: degenerate parallel transport condition} together with the assumption that $H_t$ vanishes on the orthogonal complement of the sum of the subspaces $\HH_{ja}$, and hence on the orthogonal complement of the support of $\rho_t$:
\begin{equation}
\begin{split}
    H_t^\inco 
    &= \sum_{j=1}^l\sum_{a,b=1}^{m_j}\ketbra{\psi_{ja;t}}{\psi_{ja;t}} H_t 
    \ketbra{\psi_{jb;t}}{\psi_{jb;t}} \\
    &= i \sum_{j=1}^l\sum_{a,b=1}^{m_j}\ket{\psi_{ja;t}}\braket{\psi_{ja;t}} 
    {\dot\psi_{jb;t}}\bra{\psi_{jb;t}} \\
    &= 0.   
\end{split}
\end{equation}
As a consequence, $\rho_t$ has instantaneous speed equal to the energy uncertainty (cf.~Eq.~\eqref{eq: speed}), and therefore
\begin{equation}
    \length[\rho_t] 
    = \int_0^\tau \Delta H_t \, \dt 
    = \tau \Delta E.
    \label{eq: attaatta}
\end{equation}
We next compute the energy variance:
\begin{equation}
\begin{split}
    \Delta^2H_t 
    &= \sum_{j=1}^l\sum_{a=1}^{m_j} p_j \bra{\psi_{ja;t}}H_t^2\ket{\psi_{ja;t}} \\
    &= \sum_{j=1}^l\sum_{a=1}^{m_j} p_j \braket{\dot\psi_{ja;t}}{\dot\psi_{ja;t}} \\
    &= \frac{1}{\tau^2} \sum_{j=1}^l \sum_{a=1}^{m_j} p_j\theta_{ja}(2\pi-\theta_{ja}) \\
    &= \frac{1}{\tau^2}\isobound(\bfp;U)^2.
    \label{eq: nionio}
\end{split}
\end{equation}
Equations~\eqref{eq: attaatta} and~\eqref{eq: nionio} confirm that the length of $\rho_t$ equals the isoholonomic bound associated with $U$: 
\begin{equation}
    \length[\rho_t] = \isobound(\bfp;U).
\end{equation}
It remains to verify that the holonomy of $\rho_t$ at $W$ equals $U$.
This follows from
\begin{equation}
\begin{split}
    \Gamma[\rho_t;W]
    &= W^+W_\tau \\
    &= \sum_{j=1}^l\sum_{k=1}^l\sum_{a=1}^{m_j} \sum_{b=1}^{m_k} \sqrt{\frac{p_k}{p_j}} \ket{ja}\braket{\psi_{ja}}{\psi_{kb;\tau}}\bra{kb} \\
    &= \sum_{j=1}^l\sum_{a=1}^{m_j} e^{i\theta_{ja}}\ket{ja}\bra{ja} \\
    &= U.
\end{split}
\end{equation}

\section{Summary}
\label{sec: Summary}
\noindent
Quantum speed limits set fundamental bounds on how fast quantum systems can evolve. While speed limits based on fidelity measures often become trivial for cyclic evolutions, isoholonomic inequalities offer a robust alternative for bounding the minimum duration of such processes. These inequalities relate the length of a closed trajectory in state space to the holonomy it generates, providing meaningful constraints even when the system returns to its initial state. 

In this paper, we derived isoholonomic inequalities and corresponding quantum speed limits for systems undergoing cyclic evolution through isodegenerate mixed states. Building on the gauge-theoretic framework underlying the geometric phase for mixed states, we defined natural notions of length and holonomy for closed curves in the space of isodegenerate density operators. At the core of this development is the isoholonomic bound---a holonomy-dependent invariant that provides a lower bound on the length of the curve. From this inequality, we established a Mandelstam--Tamm type quantum speed limit on the return time of unitarily evolving cyclic systems. In the case where the Hilbert space dimension is at least twice the rank of the state, the isoholonomic inequality becomes tight, and the associated speed limit can be saturated.


\appendix

\titleformat{\section}[block]{\bfseries\large}{Appendix \Alph{section}:}{0.7em}{}
\titleformat{\subsection}[block]{\bfseries\normalsize}{\Roman{section}.\Alph{subsection}}{0.7em}{}
\titleformat{\subsubsection}[block]{\bfseries\small}{\Roman{section}.\Alph{subsection}.\arabic{subsubsection}}{0.4em}{}

\titlespacing\section{0pt}{1.2em plus 4pt minus 2pt}{0.8em plus 2pt minus 2pt}
\titlespacing\subsection{0pt}{1em plus 4pt minus 2pt}{0.5em plus 2pt minus 2pt}
\titlespacing\subsubsection{0pt}{0.5em plus 4pt minus 2pt}{0.3em plus 2pt minus 2pt}

\section{The isoholonomic inequality for systems in pure states}%
\label{app: The isoholonomic inequality for systems in pure states}
\noindent
This appendix reproduces the proof of the isoholonomic inequality~\eqref{eq: isoholonomic inequality for pure states}, originally derived in Ref.~\cite{HoSo2023c}.

Let $\rho_t$ be a closed curve of pure states with minimal length among all such curves having holonomy $e^{i\theta}$, where $\theta \in [0,2\pi)$. Reparameterize $\rho_t$ proportionally to arc length, preserving its orientation, so that it returns to $\rho_0$ at $\tau = 1$. This reparameterization leaves both the length and the holonomy unchanged. We will prove that
\begin{equation}
    \lengthFS[\rho_t]^2 \ge \theta(2\pi - \theta).
\end{equation}
Since the case $\theta = 0$ is trivial, we henceforth assume $\theta > 0$.

Choose a unit vector $\ket{\psi_0}$ that projects onto $\rho_0$, and let $\ket{\psi_t}$ be the horizontal lift of $\rho_t$ starting at $\ket{\psi_0}$. Since the Hopf bundle is a Riemannian submersion (see Example~\ref{rmk: Hopf bundle}), the lift has the same constant speed as $\rho_t$. Moreover, because $\rho_t$ is a length-minimizing closed curve at $\rho_0$ with holonomy $e^{i\theta}$, its lift minimizes the length among all horizontal curves connecting $\ket{\psi_0}$ to $e^{i\theta}\ket{\psi_0}$. It follows that $\ket{\psi_t}$ is an extremal of the augmented kinetic energy functional
\begin{equation}
    \energy\big[\ket{\phi_t},\lambda_t\big]
    = \frac{1}{2} \int_0^1 \bigg( \braket{\dot\phi_t}{\dot\phi_t} + 2i\lambda_t\braket{\phi_t}{\dot\phi_t} \bigg) \dt,
\end{equation}
defined on the space of curves on the unit sphere in $\HH$ connecting $\ket{\psi_0}$ to $e^{i\theta}\ket{\psi_0}$ over the interval $[0,1]$. Here $\lambda_t$ is a Lagrange multiplier enforcing the horizontality constraint.

Every vector field along $\ket{\psi_t}$ that vanishes at the endpoints and remains tangent to the unit sphere in $\HH$ can be written in the form $-iB_t\ket{\psi_t}$, where $B_t$ is a curve of Hermitian operators satisfying $B_0 = B_1 = 0$. A corresponding variation of $\ket{\psi_t}$ is given by $\ket{\psi_{\epsilon,t}} = U_{\epsilon,t} \ket{\psi_t}$, where $U_{\epsilon,t}$ is the 
backward time-ordered exponential
\begin{equation}
    U_{\epsilon,t} = \mathcal{\tilde T} \exp\bigg(-i\epsilon\int_0^t \dot B_{s}\,\ds\bigg).
\end{equation}

Integration by parts shows that the first variation of the augmented kinetic energy is given by
\begin{widetext}
\begin{ceqn}
\begin{equation}
   \frac{\d}{\d \epsilon} \energy \big[ \ket{\psi_{\epsilon,t}},\lambda_t \big]\Big|_{\epsilon=0}
    = \frac{1}{2} \int_0^1 \tr \Big( B_t \frac{\d}{\d t} \big(i\ketbra{\psi_t}{\dot\psi_t} \\ - i\ketbra{\dot\psi_t}{\psi_t} - 2\lambda_t\ketbra{\psi_t}{\psi_t} \big) \Big) \dt.
\label{partialintegration}
\end{equation}
\end{ceqn}
\end{widetext}
Since this expression must vanish for all admissible $B_t$, we obtain the Euler--Lagrange equation for $\ket{\psi_t}$:
\begin{equation}
    \frac{\d}{\d t} \big( i\ketbra{\psi_t}{\dot\psi_t} - i\ketbra{\dot\psi_t}{\psi_t} - 2\lambda_t\ketbra{\psi_t}{\psi_t} \big) = 0.
    \label{eq: Lagrange equation}
\end{equation}
Hence the Hermitian operator
\begin{equation}\label{eq: A}
    A = i\ketbra{\dot\psi_t}{\psi_t} - i\ketbra{\psi_t}{\dot\psi_t} + 2\lambda_t\ketbra{\psi_t}{\psi_t} 
\end{equation}
is independent of time. This observation has several important consequences (see Ref.~\cite{HoSo2023c} for details):
\begin{enumerate}
    \item The Lagrange multiplier is constant: $\lambda_t = \lambda$.
    \item The support of $A$ is two-dimensional, spanned by $\ket{\psi_0}$ and $\ket{\dot\psi_0}$.
    \item The evolution of the horizontal lift satisfies the Schr\"odinger type equation $\ket{\dot\psi_t} = -i(A - 2\lambda)\ket{\psi_t}$.
\end{enumerate}
It follows that $\ket{\psi_t} = e^{-it(A-2\lambda)} \ket{\psi_0}$, and that $\ket{\psi_t}$ remains confined to the two-dimensional support of $A$. 

The eigenvalues of the restriction of $A-2\lambda$ to the support of $A$ are $a_{\pm} = -\lambda \pm ( \lambda^2 + \braket{\dot\psi_0}{\dot\psi_0} )^{1/2}$. The holonomy condition $\ket{\psi_1} = e^{i\theta}\ket{\psi_0}$ then implies $a_{\pm} = 2\pi k_{\pm} - \theta$ for some integers $k_+ > 0$ and $k_- \le 0$. Consequently, the Fubini--Study length of $\rho_t$ satisfies
\begin{equation} 
\begin{split}
    \lengthFS[\rho_t]^2
    &= \braket{\dot\psi_0}{\dot\psi_0} \\
    &= - a_- a_+ \\
    &= (\theta - 2\pi k_-) (2\pi k_+ - \theta) \\
    &\ge \theta(2\pi - \theta).
    \label{1D length estimate}
\end{split}
\end{equation}

\section{The connection form}
\label{app: The connection form}
\noindent
In this appendix, we derive the expression in Eq.~\eqref{eq: Ehresmann connection} for the connection form $\AA$. Let $W$ be an amplitude and $\dot W$ a tangent vector at $W$.
The connection form is implicitly defined by the requirement that $W\AA(\dot W)$ equals the orthogonal projection of $\dot W$ onto the vertical space at $W$.
Equivalently, $\AA$ satisfies~
\begin{equation}
    G(\dot W-W\AA(\dot W),WX)=0
    \label{eq: A1}
\end{equation}
for all skew-Hermitian operators $X$ that commute with each projection operator $\Lambda_j$. We have that
\begin{equation}
\begin{split}
    &G\big(\dot W-W\AA(\dot W),WX\big) \\
	&= \frac{1}{2}\tr\big((2W^\dagger W\AA(\dot W)+\dot W^\dagger W-W^\dagger\dot W)X\big) \\
	&= \frac{1}{2}\sum_{j=1}^l \tr(\Lambda_j(\dot W^\dagger W-W^\dagger\dot W+2W^\dagger W\AA(\dot W))\Lambda_jX).
\end{split}
\end{equation}
In the first step, we used that $W^\dagger W$ and $\AA(\dot W)$ commute; in the second, we used that $X$ commutes with the spectral projectors $\Lambda_j$, which collectively sum to the identity on $\KK$.
Since $X$ is an arbitrary skew-Hermitian operator that commutes with all the $\Lambda_j$s, the vanishing of the trace for all such $X$ implies that
\begin{equation}
    \Lambda_j(\dot W^\dagger W-W^\dagger\dot W+2W^\dagger W\AA(\dot W))\Lambda_j=0.
    \label{eq: A3}
\end{equation}
Noting that $W^\dagger W$ is invertible and commutes with each $\Lambda_j$ along with its inverse, we then obtain
\begin{equation}
    \Lambda_j \AA(\dot W) \Lambda_j 
    = \frac{1}{2}\Lambda_j(W^+\dot W-(W^+\dot W)^\dagger)\Lambda_j,
    \label{eq: A4}
\end{equation}
where $W^+$ denotes the pseudoinverse of $W$; see Eq.~\eqref{eq: Moore--Penrose inverse}.
This establishes that $\AA$ is given by the expression in Eq.~\eqref{eq: Ehresmann connection}.

If $\dot{W}$ is tangent to $\WW(\bfp,\bfL)$, where $\bfp$ is the eigenvalue spectrum of $WW^\dagger$, then $\dot W^\dagger W=-W^\dagger\dot W$. Substituting this into Eq.~\eqref{eq: A3} gives
\begin{equation}
    2\Lambda_j(W^\dagger W\AA(\dot W)-W^\dagger\dot W)\Lambda_j=0,
\end{equation}
which immediately yields the simplified expression
\begin{equation}
    \Lambda_j\AA(\dot W)\Lambda_j 
    = \Lambda_jW^+\dot W\Lambda_j 
    = p_j^{-1}\Lambda_jW^\dagger \dot W\Lambda_j.
\end{equation}
Hence, for the bundle over $\DD(\bfp,\bfm)$, the connection form reduces to the expression given in Eq.~\eqref{eq: connection in isospectral case}.

\section{Independence of the auxiliary space and the eigenprojector spectrum}%
\label{app: Independence of the eigenprojector spectrum}
\noindent
Here we prove that the isomorphism class of the bundle with connection $(\eta,\AA)$, as well as the metric 
$g$ on $\DD(\bfm)$, is independent of the choice of auxiliary Hilbert space $\KK$ and of the eigenprojector spectrum $\bfL$.

Suppose $\KK'$ is another auxiliary space and $\bfL'$ is an eigenprojector spectrum for density operators on $\KK'$ compatible with the degeneracy spectrum $\bfm$. Let $\eta'$ denote the corresponding amplitude bundle over $\DD(\bfm)$ constructed from $\KK'$ and $\bfL'$.  We will show that there exists an isomorphism of principal bundles from $\eta$ to $\eta'$ that preserves the horizontal distributions determined by the connection forms $\AA$ and $\AA'$ associated with each bundle.

Since both eigenprojector spectra are compatible with $\bfm$, there exists a unitary isomorphism $V:\KK'\to\KK$ such that $\Lambda_j' = V^\dagger \Lambda_j V$. This unitary induces a Lie group isomorphism $\kappa_{V^\dagger}:\UU(\KK;\bfL)\to\UU(\KK';\bfL')$ via $\kappa_{V^\dagger}(U) = V^\dagger U V$. Define a diffeomorphism $\Phi:\WW(\bfL)\to\WW(\bfL')$ by $\Phi(W) = WV$. This map is an equivariant isomorphism of bundles with the gauge groups identified through $\kappa_{V^\dagger}$. Indeed, for any $W$ in $\WW(\bfL)$ and $U$ in $\UU(\KK; \bfL)$, we have that $\eta'(\Phi(W)) = \eta(W)$ and $\Phi(W U) = \Phi(W)\kappa_{V^\dagger}(U)$. Hence $\Phi$ is a principal bundle isomorphism.
Equivalently, this can be expressed by stating that the following diagram is commutative:
\begin{equation*}
    \begin{tikzcd}
    \WW(\bfL)\times\UU(\KK;\bfL) \arrow[r] \arrow[dd, "\Phi\times\kappa_{V^\dagger}"'] & \WW(\bfL) \arrow[dd, "\Phi"] \arrow[dr,"\eta"] & \\
     & & \DD(\bfm) \\
    \WW(\bfL')\times\UU(\KK';\bfL')\arrow[r] & \WW(\bfL') \arrow[ur,"\eta'"']& 
    \end{tikzcd}   
\end{equation*}

The connection forms $\AA$ and $\AA'$ are related by $\Phi^*\AA'=\Ad_{V^\dagger}\circ \AA$, where $\Ad_{V^\dagger}$ denotes the Lie algebra isomorphism from $\uu(\KK;\bfL)$ to $\uu(\KK';\bfL')$ induced by conjugation with $V^\dagger$. Equivalently, for each $W$ in $\WW(\bfL)$, the diagram
\begin{equation*}
    \begin{tikzcd}
    \T_W\WW(\bfL) \arrow[r,"\AA"] \arrow[d, "d\Phi"'] & \uu(\KK;\bfL) \arrow[d, "\Ad_{V^\dagger}"] \\
    \T_{WV}\WW(\bfL')\arrow[r,"\AA'"] & \uu(\KK';\bfL')
    \end{tikzcd}   
\end{equation*}
commutes. This expresses precisely that the bundle isomorphism $\Phi$ intertwines the horizontal distributions determined by $\AA$ and $\AA'$.
The verification is straightforward: for any tangent vector $\dot{W}$ at $W$, we have that
\begin{equation}
\begin{split}
    &\AA'(d\Phi(\dot W)) \\
    &= \sum_{j=1}^l \Lambda_j'\big((WV)^+(\dot WV)-((WV)^+(\dot WV))^\dagger\big) \Lambda_j' \\
    &= \sum_{j=1}^l \Lambda_j'V^\dagger \big(W^+\dot W-(W^+\dot W)^\dagger\big) V \Lambda_j' \\
    &= V^\dagger\sum_{j=1}^l \Lambda_j \big(W^+\dot W-(W^+\dot W)^\dagger\big) \Lambda_jV \\
    &= \Ad_{V^\dagger}(\AA(\dot W)).
\end{split}
\end{equation}
where we used $(WV)^+=V^\dagger W^+$ and $\Lambda_j'=V^\dagger \Lambda_j V$.

That $\Phi$ preserves the horizontal distribution also follows from it being an isometry: Let $G$ and $G'$ denote the real parts of the Hilbert-Schmidt Hermitian inner products on $\WW(\bfL)$ and $\WW(\bfL')$, respectively. Then,
\begin{equation}
\begin{split}
    &G'(d\Phi(\dot W_1),d\Phi(\dot W_2)) \\
    &= \frac{1}{2}\tr\big((\dot W_1V)^\dagger(\dot W_2V)+(\dot W_2V)^\dagger(\dot W_1V)\big) \\
    &= \frac{1}{2}\tr(\dot W_1^\dagger\dot W_2+\dot W_2^\dagger\dot W_1) \\
    &= G(\dot W_1,\dot W_2).
\end{split}
\end{equation}
Another consequence of this fact is that the projections of $G$ and $G'$ onto $\DD(\bfm)$ coincide. Thus, the induced metric on $\DD(\bfm)$ is independent of the choice of auxiliary Hilbert space and eigenprojector spectrum.

\section{Horizontal lifts and parallel-transported spectral decompositions}
\label{app: Horizontal lifts and parallel transported spectral decompositions}
\noindent
In this appendix, we show that the relation \eqref{eq: one-to-one correspondence} establishes a one-to-one correspondence between parallel-transported spectral decompositions and horizontal lifts of the curve $\rho_t$.

We begin by constructing a horizontal lift from a given parallel-transported spectral decomposition. To this end, assume that
\begin{equation}
    \rho_t = \sum_{j=1}^l \sum_{a=1}^{m_j} p_{j;t} \ketbra{\psi_{ja;t}}{\psi_{ja;t}}
\end{equation}
is a decomposition such that the vectors $\ket{\psi_{ja;t}}$ satisfy the parallel-transport condition \eqref{eq: degenerate parallel transport condition}. Fix a $\bfL$-adapted basis of $\KK$, as in Eq.~\eqref{eq: adapted basis}, and define $W_t$ as
\begin{equation}
    W_t = \sum_{j=1}^l \sum_{a=1}^{m_j} \sqrt{p_{j;t}} \ketbra{\psi_{ja;t}}{ja}.
\end{equation}
This curve lies in $\WW(\bfL)$, since $W_t^\dagger W_t$ is a density operator with eigenprojector spectrum $\bfL$:
\begin{equation}
\begin{split}
    W_t^\dagger W_t
    &= \sum_{j,k=1}^l \sum_{a=1}^{m_j} \sum_{b=1}^{m_k} \sqrt{p_{j;t}p_{k;t}}\,                  
        \ket{ja}\braket{\psi_{ja;t}}{\psi_{kb;t}}\bra{kb}  \\
    &= \sum_{j=1}^l \sum_{a=1}^{m_j} p_{j;t} \ketbra{ja}{ja}  \\
    &= \sum_{j=1}^l p_{j;t} \Lambda_j.
\end{split}
\end{equation}
Furthermore, it projects onto $\rho_t$:
\begin{equation}
\begin{split}
    W_tW_t^\dagger
    &= \sum_{j,k=1}^l \sum_{a=1}^{m_j} \sum_{b=1}^{m_k} \sqrt{p_{j;t}p_{k;t}} 
        \ket{\psi_{ja;t}}\braket{ja}{kb} \bra{\psi_{kb;t}}  \\
    &= \sum_{j=1}^l \sum_{a=1}^{m_j} p_{j;t} \ketbra{\psi_{ja;t}}{\psi_{ja;t}}  \\
    &= \rho_t.
\end{split}
\end{equation}
To show that $W_t$ is horizontal, we observe that the parallel-transport condition \eqref{eq: degenerate parallel transport condition} ensures that
\begin{equation}
\begin{split}
    &\Lambda_j W_t^+\dot W_t\Lambda_j \\
    &= \frac{1}{2} \sum_{a=1}^{m_j} \frac{\dot p_{j;t}}{p_{j;t}} \ketbra{ja}{ja} +\sum_{a,b=1}^{m_j} \ket{ja}\braket{\psi_{ja;t}}{\dot\psi_{jb;t}}\bra{jb} \\
    &= \frac{1}{2}\sum_{a=1}^{m_j}\frac{\dot p_{j;t}}{p_{j;t}} \ketbra{ja}{ja},
\end{split}
\end{equation}
which is manifestly Hermitian. From this observation it immediately follows that $W_t$ is horizontal:
\begin{equation}
\begin{split}
    \AA(\dot W_t)
    &= \frac{1}{2} \sum_{j=1}^l \Lambda_j \big( W_t^+\dot W_t - (W_t^+\dot W_t)^\dagger\big)\Lambda_j \\
    &= \frac{1}{2} \sum_{j=1}^l \big(\Lambda_j W_t^+\dot W_t\Lambda_j - (\Lambda_j W_t^+\dot W_t\Lambda_j)^\dagger\big) \\
    &= 0.
\end{split}
\end{equation}

Next, we will demonstrate how a horizontal lift yields a parallel-transported spectral decomposition. Assume that $W_t$ is a horizontal lift of $\rho_t$. Fix a $\bfL$-adapted basis in $\KK$ and define the vectors $\ket{\psi_{ja;t}}$ as in Eq.~\eqref{eq: one-to-one correspondence}. These vectors are normalized and pairwise orthogonal:
\begin{equation}
\begin{split}
    \braket{\psi_{ja;t}}{\psi_{kb;t}}
    &= \frac{\bra{ja}W_t^\dagger W_t\ket{kb}}{\sqrt{p_{j;t}p_{k;t}}}\\
    &= \sqrt{\frac{p_{k;t}}{p_{j;t}}}\braket{ja}{kb} \\
    &= \delta_{jk}\delta_{ab},
\end{split}
\end{equation}
and each of them is an eigenvector of $\rho_t$:
\begin{equation}
\begin{split}
    \rho_t\ket{\psi_{ja;t}}
    &= \frac{W_tW_t^\dagger W_t\ket{ja}}{\sqrt{p_{j;t}}} \\
    &= \sqrt{p_{j;t}}\, W_t\ket{ja} \\
    &= p_{j;t} \ket{\psi_{ja;t}}.
\end{split}
\end{equation}
\vspace{2pt}

To verify that the vectors satisfy the parallel-transport condition \eqref{eq: degenerate parallel transport condition}, we first observe that
\begin{subequations}
\begin{align}
    \braket{\psi_{ja;t}}{\dot{\psi}_{jb;t}}
    &= \bra{ka} W_t^+\dot W_t \ket{kb} - \frac{\dot p_{j;t}\delta_{ab}}{2p_{j;t}}, \label{eq: D9} \\
    \braket{\dot\psi_{ja;t}}{\psi_{jb;t}}
    &= \bra{ka} (W_t^+\dot W_t)^\dagger \ket{kb}-\frac{\dot p_{j;t}\delta_{ab}}{2p_{j;t}}.\label{eq: D10}
\end{align}
\end{subequations}
Since $W_t$ is horizontal, we have that
\begin{equation}
\bra{ka} W_t^+\dot W_t \ket{kb} = \bra{ka} (W_t^+\dot W_t)^\dagger\ket{kb}.
\end{equation}
According to Eqs.~\eqref{eq: D9} and~\eqref{eq: D10}, this  implies that 
\begin{equation}
\braket{\psi_{ja;t}}{\dot{\psi}_{jb;t}}=\braket{\dot\psi_{ja;t}}{\psi_{jb;t}}.
\end{equation}
On the other hand, the orthonormality 
$\braket{\psi_{ja;t}}{\psi_{jb;t}}=\delta_{ab}$ gives, upon differentiation, 
\begin{equation}
\braket{\psi_{ja;t}}{\dot{\psi}_{jb;t}}=-\braket{\dot\psi_{ja;t}}{\psi_{jb;t}}.
\end{equation} 
Combining these two relations, we conclude that
\begin{equation}
    \braket{\psi_{ja;t}}{\dot{\psi}_{jb;t}}=0.
\end{equation}
Hence, the parallel-transport condition \eqref{eq: degenerate parallel transport condition} is satisfied, and
\begin{equation}
    \rho_t 
    = \sum_{j=1}^l \sum_{a=1}^{m_j} p_{j;t} \ketbra{\psi_{ja;t}}{\psi_{ja;t}}
\end{equation}
is a parallel-transported spectral decomposition.
\end{document}